%% file: Archived_Paper.tex
\PassOptionsToPackage{round}{natbib}
\documentclass[12pt]{article}

\input{header}
\usepackage{booktabs}
\usepackage{rotating}
\usepackage{setspace}
\usepackage{amsmath}
\usepackage{multirow}   
\usepackage{xcolor}
\newtheorem{assumption}{Assumption}      

\newtheorem{corollary}{Corollary}
\newtheorem{condition}{Condition}



\usepackage{comment}
\usepackage[round]{natbib}

\renewcommand{\arraystretch}{.5}
\advance \topmargin by -\headheight
\advance \topmargin by -\headsep
\evensidemargin \oddsidemargin
\let\tilde=\widetilde

\usepackage[colorlinks=true, linkcolor=blue, citecolor=blue, urlcolor=blue]{hyperref}

\begin{document}

\title{Prediction Intervals for Model Averaging\footnote{First version: October 6, 2025. We thank Denis Tkachenko and seminar participants at the National University of Singapore for their helpful comments and suggestions.}}
\date{\today }
\author{Zhongjun Qu\footnote{Corresponding author. Department of Economics, Boston University; qu@bu.edu.} \and Wendun Wang\footnote{Econometric Institute, Erasmus University Rotterdam; Tinbergen Institute.} \and Xiaomeng Zhang\footnote{Econometric Institute, Erasmus University Rotterdam; Tinbergen Institute; zhang@ese.eur.nl.}}

\maketitle

\begin{abstract}
\baselineskip=19pt
A rich set of frequentist model averaging methods has been developed, but their applications have largely been limited to point prediction, as measuring prediction uncertainty in general settings remains an open problem. In this paper we propose prediction intervals for model averaging based on conformal inference. These intervals cover out-of-sample realizations of the outcome variable with a pre-specified probability, providing a way to assess predictive uncertainty beyond point prediction. The framework allows general model misspecification and applies to averaging across multiple models that can be nested, disjoint, overlapping, or any combination thereof, with weights that may depend on the estimation sample. We establish coverage guarantees under two sets of assumptions: exact finite-sample validity under exchangeability, relevant for cross-sectional data, and asymptotic validity under stationarity, relevant for time-series data. We first present a benchmark algorithm and then introduce a locally adaptive refinement and split-sample procedures that broaden applicability. The methods are illustrated with a cross-sectional application to real estate appraisal and a time-series application to equity premium forecasting.

\vspace{5mm}

\vspace{5mm}
\end{abstract}
\thispagestyle{empty}\setcounter{page}{0}\baselineskip=19pt\newpage

\section{Introduction}

Predictive modeling in the presence of model uncertainty is a central problem in econometrics and statistics. A large literature, spanning Bayesian and frequentist approaches, has developed model averaging methods to improve prediction performance when no single model is known to be the best choice. Instead of selecting a single model, these procedures combine predictions from a set of candidate models, typically with weights determined by in-sample fit, predictive performance, or prior beliefs. Bayesian Model Averaging (BMA, \cite{raftery1997bayesian}, \cite{hoeting1999bayesian}); frequentist methods such as the smoothed information criterion (SIC, \cite{buckland1997}), Mallows Model Averaging (MMA, \cite{hansen2007}), and Jackknife Model Averaging (JMA, \cite{hansen2012}); as well as more recent ensemble approaches inspired by machine learning (e.g., \cite{breiman2001random}, \citet[Chapter~8]{hastie2009elements}), are all examples of methods that aim to improve predictive performance by combining information across models and reducing the risk of relying on a single misspecified model. This paper focuses on frequentist averaging methods, but the proposed algorithms can also be applied to BMA and ensemble approaches. Forecast combination is treated as a special case of model averaging.

Despite the breadth of the model averaging literature, existing work has focused primarily on point prediction. The construction of prediction intervals, which quantify uncertainty around point predictions, has received much less attention. While Bayesian model averaging produces credible intervals by integrating over model uncertainty (\cite{raftery1997bayesian}), the frequentist coverage validity of such intervals generally requires correct model specification. To our knowledge, there is no published procedure for constructing valid prediction intervals for frequentist model averaging methods, even when averaging weights are fixed and independent of the sample. In short, the field has developed methods for estimation, but not for inference.

This paper aims to address that gap. We face two issues in obtaining valid interval predictions in a general model averaging setting. First, since all models may be misspecified, the point prediction may not converge to its true value even asymptotically. The limit generally depends on both the data-generating process and the models included in the set. One way to address this issue is to assume it away by postulating that the true model is contained in the set. However, this runs counter to the idea of model averaging, which is based on combining multiple approximating models for prediction. In empirical work, it is also common for more than one model to receive nontrivial weights. The second issue is technical: when model weights are determined endogenously, they depend on both model misspecification and the estimation sample, making it even less clear how to construct a valid prediction interval.

We propose general-purpose prediction intervals for model averaging using conformal inference framework of \cite{vovk2005}. We do not require any of the candidate models to be correctly specified and accommodate a wide range of model combinations, including nested, disjoint, overlapping, or any combination thereof, with weights that may depend on the sample in flexible ways. We establish two forms of validity: (i) finite-sample coverage under exchangeability, relevant for cross-sectional settings, and (ii) asymptotic coverage under stationarity and weak dependence, relevant for time series forecasting. In both cases, the coverage guarantees are distribution-free (that is, they do not rely on knowing or estimating the distribution of the data-generating process) and the intervals adapt automatically to the chosen model averaging scheme without further modification.

In particular, we build on the general conformal inference framework of \cite{vovk2005} and draw on more recent developments in \cite{lei2018} for regression settings. While conformal methods have been widely studied in machine learning for constructing distribution-free prediction intervals, they have not, to our knowledge, been adapted to frequentist model averaging in any published work to date. Our results fill this gap by extending conformal inference to frequentist model averaging methods, allowing for both model misspecification and dependence in the data.

We begin by proposing a basic algorithm that directly adapts the standard conformal prediction interval, such as Algorithm 1 in \cite{lei2018} for regression, to the model averaging context. Although the adaptation is straightforward, its applications are broad, and we show that the resulting method achieves valid coverage in both exchangeable and time-series settings. The procedure is general-purpose in that it applies to a wide range of model averaging schemes. In particular, we consider equal-weight combinations, the regression approach of \cite{granger1984}, SIC of \cite{buckland1997}, MMA of \cite{hansen2007}, and JMA of \cite{hansen2012}, all of which are examined in our Monte Carlo simulations and empirical applications. We then extend this basic algorithm to incorporate local adaptivity, allowing the intervals to respond to heterogeneity or serial dependence in the data and potentially improve finite-sample performance. The main advantage of adaptive intervals is that they widen when variance is high and narrow when variance is low, a feature especially relevant for financial applications. Finally, we propose split-sample versions of both procedures, which are computationally more efficient, though they may sacrifice some statistical efficiency relative to the full-sample method. We show that this approach also enables us to encompass Bayesian model averaging and ensemble methods within our framework. All algorithms apply to both exchangeable and time-series settings.

We show that, for exchangeable data, the conditions for coverage validity are mild, as they mainly rely on a symmetry condition under which the estimates and model weights are invariant to reshuffling of the data. This holds because the proposed algorithms are designed to preserve exchangeability. Factors such as the specific method for estimating model weights, model dimension, model misspecification, the number of candidate models, or whether the models are nested do not affect validity, as none of them alters exchangeability or symmetry. For time series observations, we do not require exchangeability; instead, we rely on the same symmetry condition together with stationarity, ergodicity, and the stability of the model weights and parameter estimates. We verify these conditions for linear models under each model averaging method mentioned above. As a by-product, we also establish the convergence rates of the model averaging weights in the linear time-series setting, which may be of independent interest.

We examine the finite-sample performance of the proposed methods using both exchangeable and time-series data-generating processes. For the exchangeable case, we adopt the DGP of \cite{hansen2007}, extended to include heteroskedasticity. The DGP is an infinite-order regression, so all candidate models are misspecified. The total sample size is small, with 150 observations; in the split-sample case, 75 observations are used for training, and averaging is conducted over 16 models with 2 to 17 coefficients. We find that coverage is close to the nominal level across all methods. Split-sample intervals are only slightly longer than full-sample intervals on average, while adaptive methods produce wider intervals when a conservative approach is used to estimate conditional variance. For the time-series case, we design an experiment that mimics the empirical problem of equity premium prediction with many potential predictors, using parameter values calibrated to empirical estimates. We find that the algorithms deliver coverage close to the nominal level with 100 observations, and coverage improves further when the sample size is increased to 200. In this setting, split-sample intervals are again only slightly longer than full-sample intervals on average; however, adaptive methods do not produce much wider intervals, reflecting the fact that fewer additional parameters are estimated here than in the exchangeable setting.

We illustrate the proposed algorithms with a cross-sectional application to real estate appraisal and a time-series application to equity premium prediction. To our knowledge, this is among the first empirical examinations of interval prediction in the context of model averaging.

Real estate appraisal involves estimating property values using covariates such as location, house age, and proximity to facilities. Model averaging is a natural choice when the covariates or the functional form are uncertain. In this case, prediction intervals are especially valuable: for example, a point prediction of 1 million with a 95\% interval of [0.9, 1.1] has very different implications than [0.5, 1.5], both as collateral and for potential sellers in the market. We use the dataset of Yeh and Hsu (2018), which consists of 414 housing price observations in Taipei. We consider both nested and non-nested models, a total of 63 models in all, and use a leave-one-out procedure to evaluate coverage. The results show that all methods achieve adequate coverage. Additionally, the weighting favors estimated weights over equal weights, as larger models generate more accurate forecasts.

We then turn to a time-series application, where we use 20 variables from \cite{goyal2024} as potential predictors of the equity premium. Following common practice, each variable enters a bivariate regression, leading to 20 non-nested models for averaging. In this case, we find that using equal weights produces more accurate forecasts than estimating weights. At the same time, all methods deliver coverage close to the nominal level despite the small sample size.

The rest of the paper is organized as follows. Section 2 introduces the issue of interest: measuring prediction uncertainty under model averaging. Section 3 proposes an algorithm for constructing interval predictions for general model averaging schemes, adapting Algorithm 1 of \cite{lei2018} to the model averaging setting. Section 4 establishes formal coverage guarantees under two sets of assumptions: exact finite-sample validity under exchangeability and asymptotic validity under stationarity. It also discusses applications to a range of model averaging schemes. Section 5 proposes extensions of the basic algorithm to capture heteroskedasticity or serial correlation in the data, while Section 6 provides further extensions based on sample splitting, which also includes applications to BMA and ensemble methods. Section 7 presents Monte Carlo experiments to examine coverage properties in finite samples. Section 8 considers two empirical applications, and Section 9 concludes. 

\section{The issue of interest}

The objective of the paper is to measure prediction uncertainty under model averaging. The setting is general and is as follows. We consider averaging over \( M \) candidate models, indexed by $m=1,\ldots,M$, where the $m$th model is given by
\[
y_i = f^{(m)}(\bx_i; \boldsymbol{\theta}_m) + e_{i}, \quad i = 1, \ldots, n,
\]
with $n$ denoting the sample size; $y_i$ is a scalar random variable, and $\bx_i$ is a random vector.

Different models typically contain different explanatory variables (covariates), and \( \bx_i \) denotes the union of all variables that appear in at least one model. These models may be disjoint (sharing no common covariates), nested (where one model is a special case of another), or overlapping (sharing some but not all covariates). The functional forms can differ and may be linear or nonlinear.  None of them is assumed to encompass the true data-generating process. Each model \textit{acts as if} \( \mathbb{E}(e_i|\bx_i) = 0 \), so that \( f^{(m)}(\bx_i; \boldsymbol{\theta}_m) \) represents the mean-square optimal prediction under model \( m \). This condition need not hold for the true data-generating mechanism, allowing all models to be misspecified. The dimension of $\boldsymbol{\theta}_m$ generally differs across models. 

Although we use $i$ to index observations, the proposed algorithms also apply to time series regressions. In that context, $ y_i$ and $ x_i$  are interpreted as $ y_t$ and $ x_{t-1}$, respectively, and the $m$th estimated model is given by \[
y_t = f^{(m)}(\bx_{t-1}; \boldsymbol{\theta}_m) + e_{t}, \quad t= 1, \ldots, T. 
\] All statements in the previous paragraph apply to the time series case.

Point prediction with frequentist model averaging typically proceeds as follows. Using the data sample, each model is estimated and used to generate a predicted value for \( y_{n+1} \) at a new covariate vector \( \bx_{n+1} \), given by
\begin{equation}
f^{(m)}(\bx_{n+1}; \widehat{\boldsymbol{\theta}}_m), \quad \text{for } m = 1, \ldots, M.
\label{eq:model_predictions}
\end{equation}
This new covariate vector $\bx_{n+1}$ may represent predictors observed at the end of a time window in a time-series application or predictors for a new observation in a cross-sectional setting. With model weights \( \widehat\bw = (\widehat{w}_1, \ldots, \widehat{w}_M)^\top \), the model-averaged prediction at \( \bx_{n+1} \)  is
\begin{equation}
\widehat{\mu}_{n+1}= \sum_{m=1}^M \widehat{w}_m f^{(m)}(\bx_{n+1}; \widehat{\boldsymbol{\theta}}_m).
\label{eq:overall_predictions}
\end{equation}
The weights \( \widehat\bw \) may be fixed (e.g., equal weights) or data-driven; the choice of weights will be discussed later in the paper. The proposed algorithms allow for both cases. We assume all weights are finite and do not assume they must sum up to one. These procedures yield point predictions. Bayesian model averaging will be treated separately in Subsection 6.1.

The issue of interest, which is also the paper's main contribution, is to construct valid prediction intervals that measure prediction uncertainty at any user-specified significance level \( \alpha \in (0,1) \). Specifically, we construct an interval that contains the true out-of-sample value of \( y \) when \( \bx = \bx_{n+1} \), denoted \( y_{n+1} \), with the following coverage property:
\[
\mathbb{P}\left\{y_{n+1} \in \mathcal{C}(\bx_{n+1})\right\} \geq 1 - \alpha,
\]
which holds in finite samples when the observations are exchangeable (with i.i.d. as a special case) and holds asymptotically when the data are continuous, stationary, and ergodic time series. The coverage probability converges to the nominal level $(1 - \alpha)$ in both cases for continuous outcomes and is therefore non-conservative asymptotically.

\section{A basic algorithm for interval prediction}
	
We need to address two issues in obtaining a valid interval prediction in a general model averaging setting. First, since all models may be misspecified, the point prediction may not converge to its actual value even asymptotically. The limit generally depends on both the data generating process and the models allowed in the model set. One way to address this issue is to assume away the problem; that is, to assume that the model set contains the true model. However, this assumption runs counter to the spirit of model averaging, which rests on the idea that multiple approximating models should be combined for prediction. Moreover, in empirical applications, it is common for more than one model to receive nontrivial weights. Obtaining valid prediction intervals in the presence of general model misspecification has remained an unsolved problem. The second challenge is technical in nature: when model weights are determined endogenously, they depend on both model misspecification and the estimation sample, making it even less clear how to construct a valid prediction interval.

We use the conformal inference framework of  \cite{vovk2005} to construct a prediction interval. A  key insight from conformal inference, which we carry into our setting, is that a plausible value for the unobserved \( y_{n+1} \) should be one that conforms to the observed data sample. In particular, when we attempt to predict \( y_{n+1} \), it should yield a prediction error that is consistent with the distribution of historical prediction errors, which are computable from the data and model. Specifically, given a candidate value \( y \) for \( y_{n+1} \), we compute the implied prediction error \( \widehat{e}_{n+1} = y - \widehat{\mu}_{n+1} \) as in (\ref{eq:overall_predictions}), and compare it to the empirical distribution of in-sample prediction errors \( \widehat{e}_{i}= y_i - \widehat{\mu}_{i} \), for \( i = 1, \ldots, n \). If \( \widehat{e}_{n+1} \) is implausibly large relative to the historical errors at a chosen coverage level, then \( y_{n+1}=y \) is considered a implausible value and is excluded from the prediction interval; otherwise,  \(y \) is included in the interval. The collection of such values forms a valid prediction interval at the chosen coverage level. This powerful idea has generated a large and expanding literature for conducting distributional inference on predictions, and in this paper we adapt it to frequentist model averaging and prove its validity in general settings. 

This approach does not require any model to be correctly specified, as it relies on the notion of conformity rather than approximations around a true model. A key contribution of this paper is to show that it applies to both cross-sectional and time-series model averaging. The method remains valid as long as prediction errors, both in-sample and for the new observation, can be treated symmetrically. This symmetry condition underlies the conformal inference framework and is made precise for model averaging in the algorithm and assumptions presented below.

Our proposed algorithm involves parameter and model weights estimation using augmented samples and can be described as follows. Recall that the observed sample is given by \(\{(\bx_{i}, y_i)\}_{i=1}^n\), where $i$ denotes either a cross-sectional unit or a time index. Let \(y\) be a candidate value for \(y_{n+1}\). Then an augmented sample is given by \(\{(\bx_{i}, y_i)\}_{i=1}^{n} \cup \{(\bx_{n+1}, y)\}\). For each model, we use the augmented sample to estimate $\boldsymbol{\theta}_m$ to obtain $\widehat{\boldsymbol{\theta}}_m$, where the estimate methods can differ across models, for example, using MLE for some models, while (nonlinear) least squares for others. Similarly,  the model weights are also determined based on the augmented sample; the specific steps depend on the model averaging method and are described in Section 4.3. Next, compute the point predictions and associated prediction errors as, respectively,
\begin{align}
\mathcal{A}: \quad
\widehat{\mu}_{y,i} &= \sum_{m=1}^M \widehat{w}_m f^{(m)}(\bx_{i}; \widehat{\boldsymbol{\theta}}_m)
\quad \text{for }  i = 1, \dots, n, n+1,
\label{model averaging prediction}
\end{align}
and 
\begin{align}
\widehat{e}_{y,i} &= y_{i} - \widehat{\mu}_{y,i}, \quad i = 1, \dots, n, \quad \widehat{e}_{y,n+1} = y - \widehat{\mu}_{y,n+1}.
\label{model averaging prediction_e}
\end{align}
After that, evaluate whether the value of $\widehat{e}_{y,n+1}$ is a plausible value in relation to the historical realizations $\widehat{e}_{y,i}$ for $i=1,...n$, using comformal inference, and keep or reject the corresponding $y$ value depending on the outcome of the comparsion. Finally, these operations are repeated for other candidate values of $y_{n+1}$ to obtain a set of predictions that form a prediction interval.

The key difference of this procedure from the standard procedure for obtaining point predictions, as in~\eqref{eq:model_predictions}, is the involvement of the candidate value \( y \). This ensures that future and historical values are treated symmetrically. Changing the value of \( y \) generally affects both the prediction \( \widehat{\mu}_{y,i} \) and the residual \( \widehat{e}_{y,i}\) for all \( i = 1, \dots, n+1 \), which is reflected in the notation that both variables are indexed by $y$. Formally, we propose the following algorithm for both cross-sectional and time series data:
	
	\begin{algorithm}[H]
\setstretch{1.2}  
\begin{algorithmic}[1]  
    \REQUIRE 
    Data $\{(\bx_{i}, y_{i})\}_{i=1}^n$; coverage level $(1-\alpha) \in (0,1)$; model averaging algorithm $\mathcal{A}$; new observation $ \bx_{n+1}$ at which to compute the prediction; values $\mathcal{Y}_{\text{trial}} = \{y^\ast_1, y^\ast_2, \ldots\}$ for potential outcome $y_{n+1}$.
    
    \ENSURE 
    A prediction interval at $\bx_{n+1}$, namely $\mathcal{C}(\bx_{n+1})$.
    
    \FOR{$y \in \mathcal{Y}_{\text{trial}}$}
        \STATE Using the augmented sample $\{(\bx_{i}, y_i)\}_{i=1}^n \cup \{(\bx_{n+1}, y)\}$, estimate the parameters and model weights to construct the model averaging prediction rule $\mathcal{A}$ as in Equation~\eqref{model averaging prediction}.
        
        \STATE Calculate conformity scores: 
        \[
        R_{y,i} = \left| y_i - \widehat{\mu}_{y,i}\right| \quad \text{for } i = 1, \ldots, n, \quad \text{and} \quad R_{y,n+1} = \left| y - \widehat{\mu}_{y,n+1}\right|.
        \]
        
        \STATE Compute $p$-value of $R_{y,n+1}$: 
        \[
        \pi(y) = \frac{1 + \sum_{i=1}^{n} \mathbb{I}(R_{y,i} \geq R_{y,n+1})}{n+1}.
        \]
    \ENDFOR
    
    \RETURN 
  \[
\mathcal{C}(\bx_{n+1}) = \bigl\{\, y \in \mathcal{Y}_{\text{trial}} : \pi(y) > \alpha \,\bigr\}.
\]

\end{algorithmic}
\caption{Prediction Interval for Model Averaging}
\label{algorithm 1}
\end{algorithm}

Model averaging enters the algorithm through the first part of the for loop, which involves parameter estimation, model weight estimation, and the calculation of model predictions. The subsequent steps (i.e., parts 3 and 4) are identical to those in standard conformal inference. In particular, Algorithm 1 in \cite{lei2018} contains parts 3 and 4 for general regression prediction problems. Thus, this algorithm can be viewed as a direct adaptation of conformal inference to the model averaging setting.

Parts 4 and 6 evaluate the \(p\)-value of the new prediction residual relative to the empirical distribution of all residuals. Because a small \(p\)-value indicates that the candidate \(y\) is implausible, the algorithm includes \(y\) in the prediction set only if its \(p\)-value is strictly greater than \(\alpha\); see \cite{vovk2005} for discussions of this point. This is equivalent to comparing the new residual to the empirical \((1-\alpha)\) quantile with conservative tie handling: \(y\) is included if \(R_{y,n+1} \le R_{y,(k)}\), where \(R_{y,(k)}\) is the \(k\)-th smallest residual of all residuals and \(k=\lceil(1-\alpha)(n+1)\rceil\). No estimation of nuisance parameters is required.

The model weights are calculated using the augmented sample: this is important to ensure that the new observation ${(\bx_{n+1},y)}$ affects the estimation in the same way as any existing observation in the sample. This symmetry implies that the properties of the new prediction residual should behave like the other residuals, thus providing the basis for conformity. These weights need to be recomputed for each candidate value of $y_{n+1}$. Because these computations are independent across values and therefore fully parallelizable, we conjecture that the computation is feasible for many applications in economics and finance. Nevertheless, to accommodate applications where this computation turns out to be too costly, we will in Section 6 provide algorithms that involve sample splitting, which computes the model weights only once for the entire procedure.

The algorithm is generic in that it applies to a wide range of model averaging methods. Differences across methods affect only the choice of weights, and hence the calculation of prediction residuals. We will show that the algorithm remains asymptotically valid for time series data without further modification.

\section{Coverage validity of the algorithm}

In this section, we establish coverage validity under two sets of assumptions: exact finite-sample validity under exchangeability (with i.i.d.\ data as a special case), relevant for cross-sectional settings, and asymptotic validity under stationarity, relevant for time-series predictions.

\subsection{Finite-sample validity under exchangeability}

The assumptions we need are standard assumptions in the conformal prediction literature and are nonrestrictive in the context of model averaging. 

\begin{assumption}\label{assump:exchangeable}
The random vectors $(\bx_{1}, y_1),\ldots,(\bx_{n}, y_n), (\bx_{n+1}, y_{n+1})$ are exchangeable.
\end{assumption}

Note that a sequence of random vectors is exchangeable if its joint distribution is invariant under permutations of the indices. In other words, the ordering of the observations does not matter. Formally, a sequence \( (V_1, V_2, \ldots, V_{n+1}) \) is exchangeable if, for any permutation \( \pi \) of \( \{1,2,\ldots,n+1\} \),
\(
(V_1, V_2, \ldots, V_{n+1}) \overset{d}{=} (V_{\pi(1)}, V_{\pi(2)}, \ldots, V_{\pi(n+1)}),
\)
where \( \overset{d}{=} \) denotes equality in distribution. Every i.i.d.\ sequence is exchangeable. Conditional heteroskedasticity is permitted: the variance of \(y_i\) given \(\bx_i\) may differ across observations. Exchangeability is uncommon in time-series settings, since temporal ordering typically matters and permuting time indices alters the dependence structure. A notable exception is equicorrelated Gaussian vectors, whose symmetry makes the joint distribution permutation invariant. This assumption will not be required for time-series model averaging.

For any augmented sample \( \{(\bx_{i}, y_i)\}_{i=1}^n \cup \{(\bx_{n+1}, y)\} \) in Algorithm~1, let 
\[ \widehat{\boldsymbol{\theta}} = (\widehat{\boldsymbol{\theta}}_1, \widehat{\boldsymbol{\theta}}_2, \ldots, \widehat{\boldsymbol{\theta}}_M)
\]
denote the collection of parameter estimates from the individuals models, and let 
\[ \widehat{\mathbf{w}} = (\widehat{w}_1, \widehat{w}_2, \ldots, \widehat{w}_M) \]
denote the estimated model weights. Their dependence on $y$ is omitted for simplicity.

\begin{assumption}\label{assump:perm_invariance}
The parameter estimates $\widehat{\boldsymbol{\theta}}$ and model weights $\widehat{\mathbf{w}}$ are invariant to the permutation of the augmented sample \( \{(\bx_{i}, y_i)\}_{i=1}^n \cup \{(\bx_{n+1}, y)\} \).
\end{assumption}

In this assumption, the elements of each vector \((\bx_{i}, y_i)\) are permuted together as a set. The assumption ensures that reshuffling the sample does not affect the estimates. As a result, the fitted residuals as in (\ref{model averaging prediction_e}) inherit the exchangeability of the original data. Most estimation methods for cross-sectional data satisfy this condition. In particular, weighted least squares satisfies it as long as weights are given by the same function applied to each observation’s own data, which is typically the case. Time-series multiple regression, where \(y_i\) is regressed on \(\bx_i\) via OLS, also satisfies this assumption.  A counterexample is an estimation procedure that assigns zero weight to the observation \( (\bx_{n+1}, y) \) only, meaning that this observation is excluded from the estimation. In that case, the new observation is treated differently from the others, and its prediction residual may have different properties. As a result, the residuals are no longer exchangeable. This assumption is not tied to any particular estimation method or specific model averaging scheme. It allows users to apply the procedure with any method, including newly proposed ones, by simply verifying whether the assumption holds in their specific setting.

Under these two assumptions, Algorithm~\ref{algorithm 1} provides correct coverage in finite samples. It applies to models with either discrete or continuous outcomes, in each case allowing for discrete and continuous explanatory variables.

\begin{theorem} \label{thm:conformal_validity_exchangeable}
Suppose Assumptions~\ref{assump:exchangeable} and~\ref{assump:perm_invariance} hold. Then
\[
\mathbb{P}\left\{y_{n+1}\in\mathcal{C}(\bx_{n+1})\right\} \geq 1 - \alpha.
\]
If, in addition, for all \( y \in \mathbb{R} \) the fitted absolute residuals \( \{R_{y,i}\}_{i=1}^{n+1} \) have a continuous joint distribution, then it also holds that
\[
\mathbb{P}\left\{y_{n+1} \in \mathcal{C}(\bx_{n+1})\right\} \leq 1 - \alpha + \frac{1}{n+1}.
\]
\end{theorem}

The first inequality reflects a standard property of conformal inference; see Vovk et al.\ (2005). The second is established in Theorem~1 of \cite{lei2018} for regression settings. Both results hold here because Algorithm~\ref{algorithm 1} is constructed to ensure that prediction residuals under model averaging are exchangeable. A proof is provided in the appendix, focusing on how model averaging interacts with exchangeability. Taken together, the two inequalities imply that the proposed prediction interval is essentially exact: it never undercovers and its overcoverage does not exceed $1/(n+1)$ for any sample size. In practice, coverage may deviate slightly from these bounds because implementation requires searching over a finite grid, though the difference is minor and can be reduced by further refining the grid.

Since coverage validity requires only the exchangeability of the data and the symmetry of the estimation algorithm, it holds for both fixed and data-dependent model weights. The difference is that fixed and data-dependent weights lead to different distributions for \( R_{y,i} \); as a result, the corresponding prediction intervals are generally different. Similarly, model misspecification, the number of candidate models, and whether the models are nested do not affect coverage validity, as none of these alter either exchangeability or symmetry.
	
\subsection{Asymptotic validity under stationarity}
For time series model averaging, instead of exchangeability, we assume that the data are strict stationary and ergodic.

\begin{assumption}\label{assump:stationarity}
The random vectors $(\bx_{1}, y_1),\ldots,(\bx_{n}, y_n), (\bx_{n+1}, y_{n+1})$ are stationary and ergodic.
\end{assumption}
Recall that a process \((\bx_i, y_i)\) is stationary if the joint distribution of \((\bx_{i_1}, y_{i_1}), \ldots, (\bx_{i_k}, y_{i_k})\) for any $k>0$ is invariant to shifts in the index. That is, for all \(h\),
    \[
    (\bx_{i_1}, y_{i_1}), \ldots, (\bx_{i_k}, y_{i_k}) \overset{d}{=} (\bx_{i_1 + h}, y_{i_1 + h}), \ldots, (\bx_{i_k + h}, y_{i_k + h}).
    \]
This strict stationarity assumption generally cannot be weakened to weak stationarity (i.e., time-invariant means, variances, and covariances), because the objective here is to cover a random variable, not a fixed parameter, with a specified probability. Even without model averaging, for a linear model given by \( y_i = \bx_i'\boldsymbol\beta + e_i \), with known \( \boldsymbol\beta \), it is still not possible to construct a valid prediction interval for \( y_{n+1} \) under weak stationarity alone, since the higher order moments of \( y_{n+1} \) may differ arbitrarily from those of past observations. No method can guarantee coverage if the underlying distribution is allowed to evolve freely over time. See also \citet[pp.~828--829]{barber2023} for a discussion of this point. Meanwhile, we conjecture that this assumption may be relaxed to allow for restricted distributional changes with appropriate modifications to the algorithm. \cite{tibshirani2019conformal}  show that a weighted version of conformal inference can be used to construct distribution-free prediction intervals when the covariate distributions (i.e., of \(\bx\)) differ between the training and test data, provided the likelihood ratio between the two distributions is known. \cite{barber2023} use weighted quantiles to introduce robustness to distributional drift and derive bounds on the coverage gap under fixed weighting schemes. The latter results quantify how coverage deteriorates under distributional shifts, though they do not prescribe how to select the weights. We leave the adaptation of such methods to model averaging for further work.

The ergodicity condition in Assumption \ref{assump:stationarity}  implies that a single long realization of the process is representative of its stationary distribution. This is often a minimal requirement for conducting inference in stationary time series settings. It covers weakly dependent processes, including mixing and near-epoch dependence, and allows for conditional variation in all moments, including conditional heteroskedasticity.

The next assumption requires that $\widehat{\boldsymbol{\theta}}$ and $\widehat{\mathbf{w}}$  converge in probability to some pseudo-true values, denoted by  $\boldsymbol{\theta}_0=(\boldsymbol{\theta}_{0,1},\ldots, \boldsymbol{\theta}_{0,M})$ and $\boldsymbol{w}_0=(\boldsymbol{w}_{0,1},\ldots, \boldsymbol{w}_{0,M})$, respectively, where all components are finite constants. These pseudo-true values generally depend on both the model set and the data-generating process.

\begin{assumption}\label{assump:convergence}
As $n \to \infty$, $\widehat{\boldsymbol{\theta}} \xrightarrow{p} \boldsymbol{\theta}_0$ and $\widehat{\mathbf{w}} \xrightarrow{p} \mathbf{w}_0
$ for some finite $\boldsymbol{\theta}_0$ and $\mathbf{w}_0$ independent of $n$. 
\end{assumption}

This assumption implies that all models included in the averaging are finite-dimensional; that is, their dimensions do not diverge with the sample size. The first part of the assumption, \( \widehat{\boldsymbol{\theta}} \xrightarrow{p} \boldsymbol{\theta}_0 \), concerns the convergence of parameters in individual candidate models, which is independent of model weights and not specific to model averaging. Conditions for ensuring this has been extensively studied for both linear and nonlinear models; see, for example, White’s work on asymptotic analysis under model misspecification in \cite{white1982,white1984,white1994}. These conditions are not restrictive.

The second part of the assumption, $\widehat{\mathbf{w}} \xrightarrow{p} \mathbf{w}_0$ concerns the asymptotic behavior of model weights, which depends on the model set and the averaging scheme. \cite{ZhangZouLiangCarroll2020} prove weight convergence when averaging linear models with i.i.d.\ data and when at least one model in the candidate set encompasses the data-generating process. Their weights minimize a generalized Mallows-type criterion, with a penalty term that diverges slowly to infinity. In addition, if all models are misspecified, their Theorem~3 implies that the estimated weights converge in probability to the minimizer of the asymptotic risk function, provided that this limiting criterion has a unique minimizer.\footnote{Their proof on p.~982 shows that \( S_n^\ast(\mathbf{w}) / R_n(\mathbf{w}) \) converges uniformly to 1 over a compact set of \( \mathbf{w} \), or equivalently, that \( S_n^\ast(\boldsymbol{w}) / n \) divided by \( R_n(\mathbf{w}) / n \) converges uniformly to 1. If \( R_n(\mathbf{w}) / n \) converges uniformly to \( R(\mathbf{w}) \) and \( R(\mathbf{w}) \) has a unique minimizer, then the convergence of the estimated weights follows.} Meanwhile, \cite{HjortClaeskens2003} study the limiting distribution of model averaging estimators when all candidate models lie in a local neighborhood of the data-generating process. They show, among other results, that the AIC weights remain random in the limit, partly because AIC is not a consistent model selection criterion. Therefore, there are practical situations where this assumption holds and also important cases where it fails. To accommodate the possibility that the weights may not converge, we will also introduce an alternative assumption to Assumption \ref{assump:convergence}, which is stated as Assumption \ref{assump:stable} later in this section. In addition, we will further examine this assumption under more primitive time series conditions when explicitly considering various model averaging schemes in Subsection 4.3.

We also need to impose regularity conditions on model residuals in a local neighborhood of the pseudo-true value specified in Assumption~\ref{assump:convergence}, as is typical in asymptotic analysis. For any given parameter value and model weights, and any $i \in \{1, \dots, n, n+1\}$, define \begin{align*}
\widetilde{\mu}_{i}(\mathbf{w},\boldsymbol{\theta}) &= \sum_{m=1}^M w_{m} f^{(m)}(\bx_{i};\boldsymbol{\theta}_{m}), \\
\widetilde{e}_{i}(\mathbf{w},\boldsymbol{\theta}) &= y_{i} - \widetilde{\mu}_{i}(\mathbf{w},\boldsymbol{\theta}).
\end{align*}

\begin{assumption}\label{assump:den}


The random variable \(\widetilde{e}_i(\mathbf{w}_0, \boldsymbol{\theta}_0)\) is integrable and has a density \(f\) that is uniformly bounded in an open neighborhood of its \((1 - \alpha)\) quantile \(s\): there exists \(\bar{f} < \infty\) such that \(f(u) \leq \bar{f}\) for all \(u\) near \(s\). In addition, its distribution function is strictly increasing at \(s\).
\end{assumption}
\begin{assumption}\label{assump:local}
Let \(\mathcal{B}(\boldsymbol{\theta}_0)\) and \(\mathcal{B}(\mathbf{w}_0)\) denote open neighborhoods of \(\boldsymbol{\theta}_0\) and \(\mathbf{w}_0\), respectively, that are independent of \(n\). Assume that \(\widetilde{e}_i(\mathbf{w}, \boldsymbol{\theta})\) is continuously differentiable in \(\mathcal{B}(\mathbf{w}_0) \times \mathcal{B}(\boldsymbol{\theta}_0)\) with probability one. For each \(i\), define
\[
Z_i := \sup_{\mathbf{w} \in \mathcal{B}(\mathbf{w}_0), \boldsymbol{\theta} \in \mathcal{B}(\boldsymbol{\theta}_0)} \left\| \frac{\partial}{\partial(\mathbf{w}, \boldsymbol{\theta})} \widetilde{e}_i(\mathbf{w}, \boldsymbol{\theta}) \right\|.
\]
Assume either:
\begin{itemize}
  \item[(i)] \( \mathbb{E}[Z_i] < \infty \), and the conditional density of \(\widetilde{e}_i(\mathbf{w}_0, \boldsymbol{\theta}_0)\) given \(Z_i\) is finite; or
  \item[(ii)] A sub-Weibull tail bound holds: there exist constants \(C_0 > 0\), \(a > 0\), and \(k > 1\) such that for all \(i\),
  \[
  \mathbb{P}(Z_i > t) \leq C_0 \exp(-a t^k),
  \]
  and the estimators \(\widehat{\boldsymbol{\theta}}\) and \(\widehat{\mathbf{w}}\) converge to \(\boldsymbol{\theta}_0\) and \(\mathbf{w}_0\) at rate \((\log n)^{-2/k}\) or faster.
\end{itemize}

\end{assumption}

Assumption~\ref{assump:den} is needed to ensure that the empirical distribution of \(\widetilde{e}_i(\mathbf{w}_0, \boldsymbol{\theta}_0)\) converges uniformly to its population cumulative distribution function in a neighborhood of the \((1 - \alpha)\) quantile. Assumption~\ref{assump:local},  similar to Assumption 3 in \cite{Andrews2003}, ensures that the quantile function of \(\widetilde{e}_{i}(\mathbf{w}, \boldsymbol{\theta})\) approach that of \(\widetilde{e}_{i}(\mathbf{w}_0, \boldsymbol{\theta}_0)\) as \(\mathbf{w}\) and \(\boldsymbol{\theta}\) converge to \(\mathbf{w}_0\) and \(\boldsymbol{\theta}_0\), respectively. Its part~(i) requires that the conditional density of the residual is finite, without imposing any condition on the convergence rate of the estimators. Alternatively, part~(ii) places no restriction on the density but restricts the tails of the gradient and requires a mild convergence rate of the parameter and weight estimates. It accommodates fat-tailed distributions. As shown in the appendix, this condition can be further relaxed to allow polynomial tails, provided the convergence rate is faster than some threshold below the root-\(n\) rate. We will further this assumption when considering various model averaging schemes in Subsection 4.3.

We now present an assumption to accommodate situations where the estimated model weights may not converge to fixed values. It is essentially a stability condition which requires that the sequence of fitted values from model averaging concentrate around a sequence of stationary and ergodic random variables, so that the resulting prediction errors are approximately stationary and ergodic. It is analogous to the stability condition in \cite[Assumption A2]{lei2018}, adapted to a time series model averaging setting.

\begin{assumption}\label{assump:stable}
There exists a sequence of stationary and ergodic variables \(\widetilde{\mu}_i\) independent of $n$, such that \(y_i\) admits the decomposition:
$
y_i = \widetilde{\mu}_i + \widetilde{e}_i,
$
where \(\widetilde{e}_i\) satisfies Assumption~\ref{assump:den}, and the fitted values after model averaging are stable around \(\widetilde{\mu}_i\):
\[
| \widetilde{\mu}_i(\widehat{\boldsymbol{w}}, \widehat{\boldsymbol{\theta}}) - \widetilde{\mu}_i | \leq Z_i \epsilon_n,
\]
where \(Z_i\) is a random variable that satisfies a sub-Weibull tail bound: there exist constants \(C_0 > 0\), \(a > 0\), and \(k > 1\) such that, for all \(i\),
$
\mathbb{P}(Z_i > t) \leq C_0 \exp(-a t^k),
$
and \(\epsilon_n = O_p((\log n)^{-2/k})\).
\end{assumption}

As in \cite{lei2018}, we do not assume that \( \tilde{\mu}_i \) represents the true model. Note also that the stability condition of \cite{lei2018} amounts to assuming, in the current notation, that for large enough \( n \),
$
\mathbb{P}( \max_{i \in \{1, \ldots, n+1\}} | \mu_i(\widehat{\boldsymbol{w}}, \widehat{\boldsymbol{\theta}}) - \tilde{\mu}_i | \ge \eta_n) \le \rho_n,
$
for some sequences satisfying \( \eta_n = o(1) \) and \( \rho_n = o(1) \) as \( n \to \infty \). In our setup, we introduce a dominating variable \( Z_i \) and specify a convergence rate, thereby avoiding the need to take a supremum over \( i \). Our formulation is more natural for time series variables, which are often unbounded. The original condition in \cite{lei2018} remains sufficient for our result. The next Theorem shows that Algorithm \ref{algorithm 1} is asymptotically valid. \footnote{The theorem is currently proved under the first set of assumptions. To prove it under Assumption 7 instead, only Step 1 needs to be modified: replace \( C_n \) with the supremum over \( Z_i \).}

\begin{theorem} \label{thm:conformal_validity_stationarity}
	
	If Assumptions \ref{assump:perm_invariance}-\ref{assump:local} hold or Assumptions  \ref{assump:perm_invariance}, \ref{assump:stationarity} and \ref{assump:stable} hold, then
		\[
\mathbb{P}\left\{y_{n+1}\in\mathcal{C}(\bx_{n+1})\right\} \xrightarrow{p} 1 - \alpha.
\]
	\end{theorem}
	
We prove this result by adapting the strategy used in the proof of Theorem 1 in \cite{Andrews2003}, which establishes the asymptotic validity of critical values derived from a subsampling-like method for an F-test designed to detect structural breaks near the end of the sample. In their approach, critical values are obtained by repeatedly re-estimating the model while excluding a subset of observations—a procedure that contrasts with the conformal inference approach used in Algorithm 1, which evaluates residuals using a fixed parameter estimate and based on a symmetry-based ranking.

Theoretical analyses of conformal inference for time series are far less developed than for the exchangeable case. \citet{ChernozhukovWuthrichZhu2018} make an important contribution by studying a procedure that constructs conformal scores from blocks of residuals rather than from individual observations. Our result shows that blocking is not required for asymptotic validity. This does not rule out the possibility that blocking could improve finite-sample performance by reducing serial dependence across conformal scores. Later in the paper we take a different approach to deal with such dependence and obtain algorithms that are adaptive to serial correlation and conditional heteroskedasticity.

\subsection{Applications to model averaging methods}

In this section, we show how the algorithm can be applied to various model averaging methods and examine the plausibility of the assumptions in each case. We focus on linear models, both to provide explicit expressions and because they account for most applications of model averaging in practice. All methods considered are also evaluated through Monte Carlo experiments and applied in the empirical applications.

For notation, let \(\y = (y_1, \ldots, y_n, y)^\top \in \mathbb{R}^{n+1}\) denote the observed outcomes augmented with a candidate value $y$ for \(y_{n+1}\). Let 
\(\mathbf{X} = [\bx_1^\top, \ldots, \bx_n^\top, \bx_{n+1}^\top]^\top \in \mathbb{R}^{(n+1) \times p}\) be the regressor matrix containing all  \(p\) explanatory variables that appear in any of the $M$ candidate models. 
For each model \(m = 1, \ldots, M\), let \(p_m \leq p\) be the number of regressors and define the selection matrix \(\mathbf{C}_m \in \mathbb{R}^{p \times p_m}\) so that 
\(\mathbf{X}\mathbf{C}_m \in \mathbb{R}^{(n+1) \times p_m}\) extracts the columns of \(X\) used in model \(m\). Then model \(m\) can be written as  
\begin{equation}\label{eq:model-m}
\y = \mathbf{X}\mathbf{C}_m \boldsymbol{\beta}_m + \boldsymbol{\varepsilon},
\end{equation}
with \(\boldsymbol{\beta}_m \in \mathbb{R}^{p_m}\) the coefficient vector and \(\boldsymbol{\varepsilon} \in \mathbb{R}^{n+1}\) the error term. 

All candidate models are estimated by OLS. For the \(m\)th model, denote the fitted values by  \(
\widehat{\f}^{(m)} = (\widehat{f}_1^{(m)}, \ldots, \widehat{f}_n^{(m)}, \widehat{f}_{n+1}^{(m)})^\top \in \mathbb{R}^{n+1}
\). Collecting these fitted vectors across all candidate models yields the matrix 
\[
\widehat{\F} = (\widehat{\f}^{(1)}, \ldots, \widehat{\f}^{(M)}) \in \mathbb{R}^{(n+1) \times M},
\]
whose \(i\)th row is \(
\widehat{\f}_i = (\widehat{f}_i^{(1)}, \ldots, \widehat{f}_i^{(M)}) \in \mathbb{R}^{M}
\). As before, let \(\w = (w_1, \ldots, w_M)^\top\) denote the vector of model weights.

For assumptions, since the models are estimated with OLS, Assumption~\ref{assump:perm_invariance} is automatically satisfied, and Assumptions \ref{assump:exchangeable}, \ref{assump:stationarity}, and \ref{assump:den} are standard. We therefore do not examine these further and instead focus on Assumptions~\ref{assump:convergence} and \ref{assump:local}, which we verify using primitive conditions familiar from the time series literature for each model averaging method considered below.

\subsubsection{Fixed weights}

Simple model averaging methods with fixed weights (including equal weights) often perform well relative to more sophisticated approaches, since estimating weights can substantially increase prediction bias or variance. \citet{stock2006forecasting} and \citet{timmermann2006forecast} discuss this in the context of macroeconomic forecasting, and \citet{claeskens2016} provide expressions quantifying the effect of weight estimation on forecast precision. With fixed weights, computation reduces to repeated OLS regressions and is therefore straightforward. We now turn to the assumptions.

Assumption~4 holds under mild conditions, as it only requires
\begin{equation}\label{conv_p}
\widehat{\boldsymbol{\beta}}_m \xrightarrow{p} 
\boldsymbol{\beta}_{0,m} 
= \left( \mathbb{E}\!\left[\mathbf{C}_m^\top \mathbf{x}_i \mathbf{x}_i^\top \mathbf{C}_m \right] \right)^{-1} 
  \mathbb{E}\!\left[\mathbf{C}_m^\top \mathbf{x}_i y_i \right]
\end{equation}
for each \(m \in \{1, \ldots, M\}\). This is satisfied, for example, when the regressor matrix has full column rank and satisfies weak moment restrictions that ensure a law of large numbers for time series data. Explicitly, it holds under Assumption~\ref{assump:stationarity} together with Condition \ref{cond:LLN} stated below:
\begin{condition}\label{cond:LLN} 
(i) $\mathbb{E}[\|\bx_i\|^2] < \infty$, (ii) $\mathbb{E}[\|\bx_i y_i\|] < \infty$, and (iii) $\mathbb{E}[\bx_i \bx_i^\top]$ is positive definite.
\end{condition}
The proof of (\ref{conv_p}) is straightforward and given in the appendix for completeness, where we clarify that it does not require correct model specification (see Lemma 4).  Similar conditions are used in Theorem~3.3.7 of \cite{white2001}, which establishes a law of large numbers for the OLS estimator under dependent and identically distributed observations.

Now consider Assumption~\ref{assump:local}, where \(Z_i\) only consists of derivatives with respect to \(\boldsymbol{\beta}_m\) for \(m = 1, \ldots, M\). In particular, 
\[
\frac{\partial e_i(\mathbf{w}, \boldsymbol{\beta})}{\partial \boldsymbol{\beta}_m} 
= -w_m \mathbf{C}_m^\top \bx_i,
\]
which involves no unknown parameters. Thus, the inequality in Assumption~6 is satisfied if the regressors meet the sub-Weibull tail condition specified there, including normal distribution as a special case. The only remaining requirement in that assumption is the convergence rate of $\widehat{\boldsymbol{\beta}}_m$ to \(
\boldsymbol{\beta}_{0,m}\) for any $m$. In the appendix, Lemma 5 shows that  \[\widehat{\boldsymbol{\beta}}_m-\boldsymbol{\beta}_{0,m}=O_p(n^{-1/2}),\] under Condition \ref{cond:LLN}, Assumption~\ref{assump:stationarity}, and the additional condition stated below:
\begin{condition}\label{cond:CLT} 
(i) \(\mathbb{E}[\|\bx_i \bx_i^\top\|^{\,r}] < \infty\) and \(\mathbb{E}[\|\bx_i y_i\|^{\,r}] < \infty\) for some \(r>2\);  and
(ii) \((\bx_i, y_i)\) form a strong mixing sequence with size $2r/(r-2)$. 
\end{condition}
Condition \ref{cond:CLT} permits commonly used weakly dependent processes, such as stationary finite-order vector autoregressive moving-average (VARMA) processes. In summary, the assumptions are not restrictive for applying the algorithm to model averaging methods with fixed weights in both cross-sectional and time-series settings.

\subsubsection{\citeauthor{granger1984}'s (\citeyear{granger1984}) regression approach}
\cite{granger1984} proposed selecting the model
weights \(\w\) by regressing \(\y\) on the candidate model predictions $\widehat{\F}$:
\begin{equation}\label{eq:reg-based weight}
	\y=\widehat{\F} \w+\boldsymbol{\e},
\end{equation}
where $\boldsymbol{\e}=(e_1,\ldots,e_{n+1})$ denotes a vector of residuals. \cite{granger1984} also considered including an intercept in the regression, leading to \(
	\y=w_0+\widehat{\F} \w+\boldsymbol{\e}\). This is equivalent to adding a candidate model with only an intercept as its explanatory variable, and thus still falls within the current framework.

The application of Algorithm~\ref{algorithm 1} to this approach proceeds as follows. First, run \(M\) separate regressions as in~\eqref{eq:model-m} on a sample augmented by \(y\) to construct \(\widehat{\F}\). Second, regress \(\y\) on \(\widehat{\F}\) as in~\eqref{eq:reg-based weight} to obtain the weights \(\widehat{\w}\). Third, compute the empirical quantiles of the residuals \(\widehat{\e}\) to decide whether to keep \(y\). Finally, repeat the procedure with alternative augmented samples to obtain an interval. This is simple to implement.

We next turn to the assumptions. For Assumption~\ref{assump:convergence}, the regression coefficients still satisfy  
\(
\widehat{\boldsymbol{\beta}}_m - \boldsymbol{\beta}_{0,m} = O_p(n^{-1/2})
\)  
since their estimation is the same as in the fixed-weights case. To study the convergence of \(\widehat{\w}\), let \(f^{(m)}_i\) denote the population projection of \(y_i\) onto the explanatory variables of model \(m\):  
\(
f^{(m)}_i = \mathbf{x}_i^\top \mathbf{C}_m \boldsymbol{\beta}_{0,m},
\)
and let \(\mathbf{f}_i = (f^{(1)}_i, \ldots, f^{(M)}_i)^\top\) denote the vector of projected values under the \(M\) models. Define
\(
\mathbf{w}_0 = \left( \mathbb{E}[\mathbf{f}_i \mathbf{f}_i^\top] \right)^{-1} \mathbb{E}[\mathbf{f}_i y_i].
\)
Then, Lemma 6 in the Appendix shows that  
\begin{equation}\label{eq:beta_conv}
\sqrt{n}\,(\widehat{\w} - \w_0) = O_p(1)
\end{equation}
under Assumption~\ref{assump:stationarity}, Conditions~\ref{cond:LLN} and~\ref{cond:CLT}, and the following condition:  

\begin{condition}\label{cond:fullrank} 
\(\mathbb{E}[\mathbf{f}_i \mathbf{f}_i^\top]\) is nonsingular.
\end{condition}
This condition holds as long as there is no collinearity among model predictions asymptotically. A typical violation occurs when two models are nested and the additional regressors in the larger model have zero population coefficients. Although \citeauthor{granger1984}'s (\citeyear{granger1984}) procedure is not intended for such settings, our method remains applicable provided Assumption~\ref{assump:stable} holds. The intuition behind Assumption~\ref{assump:stable} in this case is that, even under multicollinearity, the projection of $\y$ onto the column space of $\widehat{\F}$ is unique and well-defined. Consequently, the combined predictor is identified and concentrates around the population projection, which is stationary and ergodic by Assumption~\ref{assump:stationarity}. 

The last assumption to verify is the inequality in Assumption \ref{assump:local}. Because
\[
e_i(\mathbf{w}, \boldsymbol{\beta})=y_i-\sum_{m=1}^{M} w_m \, \mathbf{x}_i^\top \mathbf{C}_m \, \boldsymbol{\beta}_m,
\]
we obtain
\begin{equation}\label{eq:residual_derivatives}
\frac{\partial e_i(\mathbf{w}, \boldsymbol{\beta})}{\partial \boldsymbol{\beta}_m} = -w_m \mathbf{C}_m^\top \mathbf{x}_i
\quad \text{and} \quad
\frac{\partial e_i(\mathbf{w}, \boldsymbol{\beta})}{\partial w_m} = -\mathbf{x}_i^\top \mathbf{C}_m \, \boldsymbol{\beta}_m.
\end{equation}
Hence, the inequality in Assumption \ref{assump:local} holds if the regressors satisfy a sub-Weibull tail condition, as in the fixed-weight case. We conclude that the algorithm is valid under mild assumptions for the intended applications of this model averaging approach.

The conclusions reached for \eqref{eq:beta_conv}  and \eqref{eq:residual_derivatives} apply equally to the model averaging methods considered below. Accordingly, we do not re-examine them for those methods.

\subsubsection{Smoothed information criteria}

\cite{buckland1997} proposed defining model weights using the Akaike Information Criterion (AIC) or the Bayesian Information Criterion (BIC): $$
\widehat{w}_m=\frac{\exp(-IC_m/2)}{\sum_{s=1}^M\exp(-IC_s/2)},
$$ 
where $IC_m$ is the information criterion value for the $m$th model. For linear regression models estimated by OLS, these are
\[
\text{AIC}_m = (n+1) \log(\widehat{\sigma}_m^2) + 2p_m, \quad \text{and} \quad \text{BIC}_m = (n+1) \log(\widehat{\sigma}_m^2) + p_m \log(n+1),
\]
where \(\widehat{\sigma}_m^2\) is the estimated error variance from model \(m\), \(p_m\) is the number of parameters in model \(m\), and \(n+1\) is the augmented sample size. 

To implement Algorithm~\ref{algorithm 1} with this model averaging approach, first estimate the \(M\) models for each augmented sample and compute the corresponding information criteria to obtain the model weights. The same procedure is then repeated for other augmented samples to construct the prediction interval. These steps mirror the application of our conformal inference procedure to \citeauthor{granger1984} (\citeyear{granger1984})’s model averaging approach.

We next examine Assumptions~\ref{assump:convergence} and~\ref{assump:local}. We focus on the model weights; the remaining parts require no further discussion, as noted above. Lemma 7 in the appendix extends \citet{sin1996} and shows that, under fixed parameters and model dimensions, the estimated weights asymptotically concentrate on the model(s) that minimize the per-observation Kullback–Leibler information criterion (KLIC), with exponential or polynomial rates in the sample size determined by the KLIC gaps.

We now provide more details for this convergence result. Note that the KLIC for the $m$th model, up to an additive constant, is \[Q_m(\boldsymbol{\theta}_m)=\mathbb{E}\!\left[-\log \phi\!\left(y_i;\,\mathbf{x}_i^\top \mathbf{C}_m \boldsymbol{\beta}_m,\,\sigma^2\right)\right],
\] with $\phi(y;\mu,\sigma^2)=(2\pi\sigma^2)^{-1/2}\exp\!\left\{-(y-\mu)^2/(2\sigma^2)\right\}$. Let $Q_m$ denote $Q_m(\boldsymbol{\theta}_m)$ evaluated at its pseudo-true value $\boldsymbol{\theta}_{0,m}$. Let $m^\star \in \arg\min_{m} Q_m$ denote an index of a model that attains the smallest KLIC (when the minimizer is unique, $m^\star$ is that model). Writing the IC gap for model $m$ from $m^\star$ as $IC_{m}-IC_{m^\star} = (n+1)\,\Delta_{m}+ c_n\,(p_m - p_{m^\star}) + \varepsilon_{m,n}$, where $\Delta_{m}=Q_m - Q_{m^\star}$ is the KLIC gap, $c_n = 2$ for AIC and $c_n = \log (n+1)$ for BIC, and $\varepsilon_{m,n} = O_p(1)$. Then, under Assumption~\ref{assump:stationarity}, Conditions~\ref{cond:LLN} (without $iii$) and~\ref{cond:CLT}, Lemma 7 implies the following three results: \textbf{(1}) If $m^{\star}$ is unique, then for both AIC and BIC weights, $\widehat{w}_{m^\star}-1 = O_p\!\big(e^{-n \Delta_{\min}/2}\big)$ and $\widehat{w}_m = O_p\!\big(e^{-n \Delta_m/2}\big)$ for every $m \ne m^\star$, where $\Delta_{\min} = \min_{j\ne m^\star}\Delta_j$. Therefore, in this case the estimated model weights converge to 1 or 0 at exponential rates. \textbf{(2)} If multiple models achieve the smallest KLIC, then the BIC weights concentrate on the tied model with the fewest parameters, denoted $m_0$ with $p_{\min}$ parameters: $\widehat{w}_{m_0}-1 = O_p\!\big(n^{-d_{\min}/2}\big)$ and, for any tied model with $p_m>p_{\min}$ parameters, $\widehat{w}_m = O_p\!\big(n^{-(p_m - p_{\min})/2}\big)$, where $d_{min}>0$ is the parameter gap between the two most parsimonious tied models. For the remaining models,  $\widehat{w}_m = O_p\!\big(e^{-n \Delta_m/2}\big)$. Thus, the BIC model weights converge to 1 or 0 at polynomial rates (within the tied set) or at exponential rates (outside the tied set). \textbf{(3)}  If multiple models attain the smallest KLIC, then the AIC weights satisfy \(\sum_{m\in\mathcal{K}_{\min}} \widehat{w}_m-1
=O_p\!\big(e^{-n\Delta^\circ/2}\big)
\) and $\widehat{w}_m = O_p\!\big(e^{-n \Delta_m/2}\big)$ for $m\notin \mathcal{K}_{\min}$, where $\mathcal{K}_{\min}=\{m:\Delta_m=0\}$ and  $\Delta^\circ=\min_{m\notin\mathcal{K}_{\min}}\Delta_m>0$. Thus, the total weight on the best-performing models converges to 1, and the weights on all other models converge individually to 0 at exponential rates. However, within the best-performing set, the weights may remain random even asymptotically.

Together, the results show that Assumptions~\ref{assump:convergence} and~\ref{assump:local} hold under mild conditions, except when more than one model attain the smallest KLIC and AIC weights are used for model averaging. The theoretical properties of Algorithm~\ref{algorithm 1} in this case are left for future work.

\subsubsection{Mallows criterion}
\cite{hansen2007} proposes selecting the weights by minimizing a Mallows criterion:
\begin{eqnarray}
	\calC(\w)=\|\y-\widehat{\F}\w\|^2+2\hat{\sigma}^2 \sum_{m=1}^M w_m p_m\label{eq:formG1},
\end{eqnarray}
where \( \widehat{\sigma}^2 \) is an estimator of the error variance, typically obtained from the largest fitted model, and $\sum_{m=1}^M w_m p_m$ denotes the effective number of parameters of the averaged model, where \( p_m \) is the number of regressors in model \( m \). \cite{hansen2007} shows that the resulting weights are asymptotically mean-square optimal for independent data. \cite{hansen2008} considers its application to time series forecasting, while \cite{liao2021} establishes its optimality when the data-generating process is an infinite-order autoregression. The latter paper also establishes convergence rates for the estimated weights when the limit is unique and an interior point.

Applying Algorithm~\ref{algorithm 1} here involves running \(M\) regressions on an augmented sample, minimizing the objective function (\ref{eq:formG1}) to obtain Mallows weights, computing empirical residual quantiles, and repeating the process with alternative augmented samples. The key difference from the three model averaging approaches discussed earlier is an optimization step to compute the model weights. Since this step is a quadratic program in $M$ variables, the additional computational cost is typically small.

To verify Assumptions~\ref{assump:convergence} and \ref{assump:local}, 
Lemmas 8 and 9 in the Appendix show that
\[
\|\widehat{\mathbf w}-\mathbf w_0\| = O_p(n^{-1/2}),
\]
in two practical settings:  
(i) when a single model uniquely minimizes the population risk 
(equivalently, has the lowest per-observation KLIC, as defined in smoothed information-criterion weighting), 
in which case the estimated weights concentrate on that model, so that $\mathbf w_0$ is a unit vector with a one in the position of the best model; and  
(ii) when several models tie in KLIC, in which case convergence holds 
provided the population loss  
\[
\mathcal C_\infty(\mathbf w)
= \mathbb{E}\!\left[\Big(y_{i+1}-\sum_{m=1}^M w_m f_{i+1}^{(m)}\Big)^2\right]
\]
has a unique minimizer on the simplex (i.e., a unique “best mix” of models), so that the limiting vector $\mathbf w_0$ contains multiple nonzero entries corresponding to the optimal mix. These results hold under Conditions~\ref{cond:LLN} and~\ref{cond:CLT} and Assumption~\ref{assump:stationarity}.

These two lemmas do not cover the case where the population loss has multiple minimizers, i.e., when different model combinations yield the same population risk. In that case, the estimated weights may remain random asymptotically. Analysis of the algorithm in this setting is left for future work.

\subsubsection{Jackknife weighting}

\cite{hansen2012}  proposed a jackknife model averaging (JMA) estimator which selects the weights by minimizing a cross-validation criterion: 
\begin{align}\label{eq:formH}
	\CV(\w) = |\y - \bar{\F}\w|^2,\nonumber
\end{align}
where $\bar{\F}$ is the same as $\widehat{\F}$, except that its \( i \)th row, which corresponds to the prediction for \( y_i \), is obtained from OLS regressions excluding the \( i \)th observation, rather than using all observations as in $\widehat{\F}$. This leave-one-out construction is applied to each row of $\bar{\F}$. \cite{hansen2012} showed that the estimator is asymptotically optimal in the sense of achieving the lowest possible expected squared error for random samples, allowing for conditional heteroskedasticity. \cite{zhang2013} extended the optimality result to a limited class of time series models, including infinite-order autoregressions and models with serially correlated errors and strict exogeneity.

The application of the algorithm to JMA follows similar steps as in the Mallows case, except that it's more computationally costly due to the repeated leave-one-out estimation. The convergence properties of the weights are similar to the Mallows case as well. If the population loss has a unique minimizer on the simplex, then the jackknife model averaging weights converge to that target at the usual \(\sqrt{n}\) parametric rate under standard time-series conditions (Assumption~\ref{assump:stationarity}, Conditions~\ref{cond:LLN} and~\ref{cond:CLT}). Intuitively, the leave-one-out criterion uniformly tracks the population prediction risk $\mathcal C_\infty(\mathbf w)$, and the risk’s positive curvature around its unique minimum delivers the \(\sqrt{n}\) rate. This covers both the case where a single model is optimal (weights concentrate on that model) and the case where a unique mixture is optimal (weights concentrate on that mix). If the population loss admits multiple minimizers, one may obtain convergence only to the minimizer set unless a tie-breaking rule is imposed, again mirrors the Mallows case.

\section{Locally adaptive predictive intervals}
Algorithm \ref{algorithm 1} is simple to implement; however, it does not make use of properties of the model residuals beyond basic assumptions. As a result, the prediction interval is not adaptive and remains approximately constant across different covariate values, leaving useful information unused. For example, if the residuals are conditionally heteroskedastic, this information can be incorporated to obtain intervals that  are narrower when the conditional variance is low and wider when it is high. This issue arises for conformal prediction broadly and is not specific to model averaging. It has been examined in recent work such as \citet{lei2018}, \citet{romano2019conformalized}, and \citet{chernozhukov2021}. Building on this line of research, we develop methods tailored to the model averaging setting. The main idea is to introduce a statistical model for the model-averaging residuals and use it to prewhiten them before constructing the prediction interval.

As before, let $y$ be any candidate value for $y_{n+1}$. For simplicity, denote the model-averaging residual $\widehat e_{y,i}$ in (\ref{model averaging prediction_e}) by $\widehat e_{i}$. We introduce a statistical model for these residuals. This model is not assumed to coincide with the true data-generating process; rather, it serves as a device to capture some degree of conditional heteroskedasticity and, in the time-series setting, potential serial correlation. We treat the exchangeable and time-series cases separately, with the former building on the approach of \citet{lei2018}.


For exchangeable observations, we consider the following model for the model averaging residuals:
\[
\widehat{e}_i = \sigma_i \eta_i \quad (i = 1, \ldots, n+1),
\]
with
\[
\sigma_i^2 = \exp(\bx_i^\top \boldsymbol{\gamma}).
\]

The term \( \eta_i \) is assumed to be i.i.d.\ with mean zero and unit variance. The parameter \( \boldsymbol{\gamma} \) can be estimated by regressing \( \log(\widehat{e}_i^2) \) on \( {\x}_i \) using OLS, or by maximum likelihood under the assumption that \( \widehat{e}_i \sim \mathcal{N}(0, \exp({\x}_i^\top \boldsymbol{\gamma})) \). If only a subset of covariates is expected to influence heteroskedasticity, one may consider a subset or use a penalized regression (e.g., LASSO) to estimate \( \boldsymbol{\gamma} \). Once \( \boldsymbol{\gamma} \) is estimated, the standardized residuals are computed as
\begin{equation}
\widehat{\eta}_{y,i}:=\widehat{\eta}_i = \frac{\widehat{e}_i}{\widehat{\sigma}_i}, \quad \text{with} \quad \widehat{\sigma}_i^2 = \exp({\x}_i^\top \widehat{\boldsymbol{\gamma}}).
\label{eta_cross_section}
\end{equation}
\citet{lei2018} regress absolute residuals on covariates to estimate local variability, which can also be adopted here. By contrast, the log-linear specification guarantees positivity and accommodates both OLS and maximum likelihood estimation.

In the time-series case, we propose modeling the residuals with an AR(1)–GARCH(1,1) specification:  
\begin{equation*}
\widehat{e}_{i+1} = \delta + \rho \widehat{e}_i + \sigma_i \eta_{i+1}, 
\quad 
\sigma_i^2 = c + \alpha \widehat{e}_i^2+ \beta \sigma_{i-1}^2,
\end{equation*}
where \( i = 1, \ldots, n \), and \( \eta_{i+1} \) is assumed to be i.i.d.\ with mean zero and unit variance. The AR(1) component is in the same spirit as \citet{andrews1991}, who used it as a parsimonious approximation to more general data-generating processes to obtain improved covariance estimates in the presence of heteroskedasticity and autocorrelation of unknown forms. The GARCH(1,1) specification is motivated by empirical evidence that it is often a competitive model for capturing financial volatility; see \citet{hansen2005}. This proposal does not preclude alternative specifications—for example, replacing AR(1) with an MA(\(k\)) process in the conditional mean equation when appropriate for the application, or omitting the AR(1) component entirely if residual serial correlation is expected to be negligible.

The model is estimated by Gaussian maximum likelihood, as is standard practice. To initialize the recursion, we set
\(
\sigma_0^2 = c/(1 - \alpha - \beta),
\)
which corresponds to the unconditional variance of a stationary GARCH(1,1) process. After estimation, the conditional standard deviations \( \widehat{\sigma}_i \) are obtained recursively using the fitted values of \( (\delta, \rho, c, \alpha, \beta) \). The residuals are then standardized by first removing serial correlation and subsequently dividing by the conditional standard deviation, as
\begin{equation}
\widehat{\eta}_{y,i+1}:=\widehat{\eta}_{i+1} = \frac{\widehat{e}_{i+1} - \widehat{\delta} - \widehat{\rho} \widehat{e}_i}{\widehat{\sigma}_i}.
\label{eta_time_series}
\end{equation}


Putting the pieces together, we propose the following algorithm, which replaces the residuals \( \widehat{e}_{i} \) in Algorithm~\ref{algorithm 1} with the standardized residuals \( \widehat{\eta}_{i} \). The procedure applies to both exchangeable and time-series data.

\begin{algorithm}[H]
\setstretch{1.2}  
\begin{algorithmic}[1]  
    \REQUIRE 
    Data $\{(\bx_{i}, y_{i})\}_{i=1}^n$; coverage level $(1-\alpha) \in (0,1)$; model averaging algorithm $\mathcal{A}$; new observation $\bx_{n+1}$ at which to construct the prediction interval; values $\mathcal{Y}_{\text{trial}} = \{y^\ast_1, y^\ast_2, \ldots\}$ for potential outcome $y_{n+1}$; and a parametric model to approximate conditional heteroskedasticity or serial correlation of the prediction residuals.
    
    \ENSURE 
    Prediction interval at $\bx_{n+1}$, namely $\mathcal{C}(\bx_{n+1})$.
    
    \FOR{$y \in \mathcal{Y}_{\text{trial}}$}
        \STATE Using the augmented data $\{(\bx_{i}, y_i)\}_{it=1}^n \cup \{(\bx_{n+1}, y)\}$, compute the weights and the model averaging prediction rule $\mathcal{A}$ as in Equation~\eqref{model averaging prediction}.
        
\STATE Compute the standardized residuals $\widehat{\eta}_{y,i}$, using equation~\eqref{eta_cross_section} for the exchangeable data case and equation~\eqref{eta_time_series} for the time series case. Obtain generalized conformity score:
  \[
        R_{y,i} = \left|\widehat{\eta}_{y,i}\right|\quad \text{for } i = 1, \ldots, n, n+1.
        \]
        
        \STATE Compute p-value of $R_{y,n+1}$: 
        \[
        \pi(y) = \frac{1 + \sum_{i=1}^{n} \mathbb{I}(R_{y,i} \geq R_{y,n+1})}{n+1}.
        \]
    \ENDFOR
    
    \RETURN 
    \[
\mathcal{C}(\bx_{n+1}) = \left\{ y \in \mathcal{Y}_{\text{trial}} \,\middle|\, \pi(y) >\alpha\right\}.
\]

\end{algorithmic}
\caption{Locally Adaptive Prediction Interval for Model Averaging}
\label{algorithm 2}
\end{algorithm}

For exchangeable data, the algorithm is valid in finite samples because standardizing the residuals does not affect exchangeability, since each residual \( e_i \) is specified as a function of its own covariates \( \bx_i \) only. For time series data, where asymptotic validity does not require exchangeability, the algorithm is asymptotically valid if the standardized residuals are approximately stationary and ergodic in large samples. We formalize these results in the following corollary. Let \( \boldsymbol{\phi} = \boldsymbol{\gamma} \) in the exchangeable case, and \( \boldsymbol{\phi} = (\delta, \rho, c, \alpha, \beta) \) in the time series case.
\begin{corollary}\label{cor: adaptive}
(i) For the exchangeable data case, the conclusion of Theorem~\ref{thm:conformal_validity_exchangeable} holds under the same conditions stated there. (ii) For the time series case, if Assumptions \ref{assump:perm_invariance} and \ref{assump:stationarity} hold and Assumptions \ref{assump:convergence}-\ref{assump:local} are satisfied with $\boldsymbol \theta$ replaced by $(\boldsymbol \theta,  \boldsymbol \phi)$ and $e_i$ replaced by $\eta_i$, then 
\[
\mathbb{P}\left\{y_{n+1}\in\mathcal{C}(\bx_{n+1})\right\} \xrightarrow{p} 1 - \alpha.
\]
\end{corollary}

The assumptions in Corollary~\ref{cor: adaptive} are similar to those in Theorems~\ref{thm:conformal_validity_exchangeable} and \ref{thm:conformal_validity_stationarity}, with modifications arising from standardization. Assumptions~\ref{assump:exchangeable}--\ref{assump:stationarity} are unchanged, and Assumption~\ref{assump:den} remains standard. For Assumption~\ref{assump:convergence}, the convergence of \(\widehat{\boldsymbol{\theta}}\) and \(\widehat{\mathbf{w}}\) is unchanged, but the assumption is strengthened to also require convergence of the additional parameters \(\boldsymbol{\phi}\) to fixed values. Assumption~\ref{assump:local} is also modified: it now requires that the derivatives of the standardized residuals (rather than the original residuals) with respect to the coefficients and weights satisfy regularity conditions. Assumption~\ref{assump:stable} does not enter the corollary. Thus, relative to the earlier results, the main strengthenings are in Assumptions~\ref{assump:convergence} and \ref{assump:local}, together with the exclusion of cases covered by Assumption~\ref{assump:stable}; these changes pertain only to the time-series setting.

The proof for (i) is the same as that of Theorem \ref{thm:conformal_validity_exchangeable}. The proof for (ii) is the same as that of Theorem~\ref{thm:conformal_validity_stationarity} after replacing $\boldsymbol \theta$ and $e_i$ with $(\boldsymbol \theta,  \boldsymbol \phi)$ and $\eta_i$ when relevant. We therefore we do not include separate proofs in the appendix.

\subsection{Applications to model averaging methods}
Algorithm~\ref{algorithm 2} can be applied to all model averaging methods discussed in Subsection~4.5. Because the standardization step is applied after the model weights are computed, the additional computational cost is marginal.

Regarding the assumptions for the time series case, verifying the convergence of the AR(1)-GARCH(1,1) parameters and the properties of the residual gradients in Assumption~\ref{assump:local} is cumbersome due to the added nonlinearity and dynamics and provides little additional insight. Instead, we evaluate the procedure through simulations and empirical applications to assess the trade-offs introduced by the additional assumptions. The results suggest that standardization does not affect coverage, but may lead to noticeably wider intervals when many covariates are included in the variance model.

\section{Split-sample prediction intervals}
If computational feasibility outweighs prediction accuracy, for example when the sample is large or the models are complex, split-sample procedures are attractive. Sample splitting is common in the conformal inference literature (see, e.g., \cite{lei2018}). Below we introduce a sample-split version of Algorithm \ref{algorithm 1}.

\begin{algorithm}[H]
\setstretch{1.2}
\begin{algorithmic}[1]
    \REQUIRE 
    Data $\{(\bx_{i}, y_{i})\}_{i=1}^{n}$; coverage level $(1-\alpha) \in (0,1)$; model averaging algorithm $\mathcal{A}$; new observation $\bx_{n+1}$ at which to construct the prediction;  values $\mathcal{Y}_{\text{trial}} = \{y_1^*, y_2^*, \ldots\}$ for potential outcome $y_{n+1}$.
    
    \ENSURE 
   Prediction interval $\mathcal{C}_\mathcal{A}(\bx_{n+1})$.
    
    \STATE Split the index set $\{1, \ldots, n\}$ into two subsets $\mathcal{I}_1$ and $\mathcal{I}_2$ of (approximately) equal size.
    
    \STATE Estimate coefficients and model weights on $\mathcal{I}_1$ to obtain the prediction rule $\widehat{\mu}(\cdot)$ in $\mathcal{A}$.\footnotemark
    
    \STATE Compute conformity score for each $i \in \mathcal{I}_2$ using the estimated prediction rule $\widehat{\mu}(\cdot)$:
    \[
    R_{i}= \left| y_i - \widehat\mu(\bx_i) \right|.
    \]
    
    \FOR{$y \in \mathcal{Y}_{\text{trial}}$}
        \STATE Compute conformity score for candidate $y$:
        \[
        R_{y,n+1} = \left| y - \widehat\mu(\bx_{n+1}) \right|.
        \]
        
        \STATE Compute p-value of $R_{y,n+1}$:
        \begin{equation}
        \pi(y) = \frac{1 + \sum_{i \in \mathcal{I}_2} \mathbb{I}(R_{i}\geq R_{y,n+1})}{|\mathcal{I}_2| + 1}.
        \label{eq: p-value of split_conformal_full_style}
        \end{equation}
    \ENDFOR

    \RETURN 
    \begin{equation}
    \mathcal{C}(\bx_{n+1}) = \left\{ y \in \mathcal{Y}_{\text{trial}} \,\middle|\, \pi(y) >\alpha\right\}.
     \label{eq: interval of split_conformal_full_style}
    \end{equation}
\end{algorithmic}
\caption{Split-sample Prediction Interval for Model Averaging}
\label{algorithm:split_conformal_full_style}
\end{algorithm}
\footnotetext{Since the coefficient and weight estimates do not depend on candidate $y$ for $y_{n+1}$, $\widehat{\mu}_{y,i}$ simplifies to $\widehat{\mu}(\bx_i)$.}

For exchangeable data, the sample splitting should be random, independent of both covariates and outcomes. For time series data, the split should follow the time order, with the first half used for estimation and the second half for inference. Computationally, $R_{y,n+1} $ exceeds the ${(1-\alpha)}$ quantile of $\{R_i\}_{i\in\mathcal I_2}\cup\{R_{y,n+1}\}$ if and only if it exceeds ${(1-\alpha)}$ quantile of $\{R_i\}_{i\in\mathcal I_2}\cup\{\infty\}$ (see Lemma 1 in \cite{tibshirani2019conformal}). Therefore, the set 
\(\mathcal C(\mathbf x_{n+1})\) can be computed equivalently as  
\(
\mathcal C(\bx_{n+1}) = [\,\hat\mu(\bx_{n+1}) - d,\; \hat\mu(\bx_{n+1}) + d\,]
\), where \(d\) is the \(k\)-th smallest value of \(\{R_i: i\in\mathcal I_2\}\) with $k=\big\lceil (|\mathcal I_2|+1)(1-\alpha)\big\rceil$. This formula is used in Algorithm 2 of \cite{lei2018} to construct split conformal prediction intervals for general regression problems.

The computational saving comes from estimating model weights only once. This makes it feasible to apply otherwise costly methods, such as ensemble approaches (e.g., bootstrap aggregation, or bagging; \cite{breiman1996bagging}) and Bayesian model averaging. In bagging, the bootstrap is applied to a training subsample to obtain estimates, with the remaining sample used to construct the prediction interval. In Bayesian model averaging, the training subsample is used to estimate model weights and parameters (e.g., posterior means), and the remaining sample to construct the interval. The tradeoff is a potential loss of prediction accuracy relative to the full-sample case, possibly leading to wider intervals. We examine this tradeoff in simulations and empirical applications. The next result establishes finite-sample validity for exchangeable data and asymptotic validity for time series data.

\begin{corollary} \label{thm:conformal_validity_split}
	
(i) For exchangeable data, under Assumption~\ref{assump:exchangeable}, $\mathbb{P}\left\{y_{n+1}\in\mathcal{C}(\bx_{n+1})\right\}\geq 1-\alpha.$
		If, in addition, the fitted absolute residuals $|y_{i}-\widehat{\mu}(\bx_{i})|$, $i\in\{[n/2+1],\ldots,n+1\}$ have a continuous joint distribution for all $y\in \mathbb{R}$, then it also holds that
		$\mathbb{P}\left\{y_{n+1}\in\mathcal{C}(\bx_{n+1})\right\}\leq 1-\alpha+1/(|\mathcal{I}_2|+1).$
(ii) For time series data, if Assumptions \ref{assump:stationarity}-\ref{assump:local} hold, or if Assumptions~\ref{assump:stationarity} and \ref{assump:stable} hold, then
$
\mathbb{P}\left\{y_{n+1}\in\mathcal{C}(\bx_{n+1})\right\} \xrightarrow{p} 1 - \alpha.
$
\end{corollary}

For exchangeable data, because parameters and model weights are estimated on $ \mathcal{I}_1$, they do not interact with the exchangeability of the residuals on $ \mathcal{I}_2$, therefore Assumption \ref{assump:perm_invariance} is no longer needed. For the time-series case, since the $(n+1)$th observation does not enter the estimation, this assumption is similarly not relevant. This simplification implies that the resulting prediction interval is valid for estimation methods involving resampling, such as bagging. It's also valid for Bayesian model averaging, where the prior and estimates are obtained independent of $ \mathcal{I}_2$. More details on these two applications are given at the end of this section.  The proof of this corollary is essentially the same as that of Theorems~\ref{thm:conformal_validity_exchangeable} and~\ref{thm:conformal_validity_stationarity}, in fact simpler because the $(n+1)$th observation is independent of the estimation, and is therefore omitted.

For some applications, combining sample splitting with local adaptability is useful, as it reduces computational cost relative to full-sample estimation while still adapting to the residual structure at $n+1$. We propose the following algorithm:

\begin{algorithm}[H]
\setstretch{1.2}
\begin{algorithmic}[1]
    \REQUIRE 
    Data $\{(\bx_{i}, y_{i})\}_{i=1}^{n}$; coverage level $(1-\alpha) \in (0,1)$; model averaging algorithm $\mathcal{A}$; new observation $\bx_{n+1}$ at which to construct the prediction; values $\mathcal{Y}_{\text{trial}} = \{y_1^*, y_2^*, \ldots\}$ for potential outcome $y_{n+1}$; and a parametric model for residuals to capture conditional heteroskedasticity or serial correlation.
    
    \ENSURE 
    Prediction interval $\mathcal{C}_\mathcal{A}(\bx_{n+1})$.
    
    \STATE Split the index set $\{1, \ldots, n\}$ into two subsets $\mathcal{I}_1$ and $\mathcal{I}_2$ of (approximately) equal size.
    
    \STATE Estimate coefficients and model weights on $\mathcal{I}_1$ to obtain the prediction rule $\widehat{\mu}(\cdot)$ and residuals. Fit a parametric model to these residuals to estimate the heteroskedasticity and/or serial dependence parameters $\widehat{\phi}$.
    
    \STATE Apply $\widehat{\mu}(\cdot)$ and $\widehat{\phi}$ to $\mathcal{I}_2$ to compute the standardized residuals, using equation (\ref{eta_cross_section}) for exchangeable data and (\ref{eta_time_series}) for time series data. Calculate  the generalized conformity score for $i \in \mathcal{I}_2$:
    \(
    R_{i}= \left|\widehat{\eta}_i\right|.
    \)
    \FOR{$y \in \mathcal{Y}_{\text{trial}}$}
        \STATE Compute conformity score for candidate $y$:
        \(
        R_{y,n+1} =\left|\widehat{\eta}_{y,n+1}\right|.
        \)
        \STATE Compute p-value of  $R_{y,n+1}$:
        \[
        \pi(y) = \frac{1 + \sum_{i \in \mathcal{I}_2} \mathbb{I}(R_{i} \geq R_{y,n+1})}{|\mathcal{I}_2| + 1}.
        \]
    \ENDFOR

    \RETURN 
    \[
    \mathcal{C}(\bx_{n+1}) = \left\{ y \in \mathcal{Y}_{\text{trial}} \,\middle|\, \pi(y) >\alpha \right\}.
    \]
\end{algorithmic}
\caption{Locally Adaptive Prediction Intervals with Sample Splitting for Model Averaging}
\label{algorithm:split_conformal_adaptive}
\end{algorithm}

As in Algorithm \ref{algorithm:split_conformal_full_style}, the model weights are computed only once. All model coefficients, including those capturing heteroskedasticity and serial correlation, are estimated from the first subsample. Also, the prediction interval can be equivalently computed as	$\mathcal{C}(\bx_{n+1}) =  [ \hat\mu(\x_{n+1}) - d\hat\sigma_{n+1},  \hat\mu(\x_{n+1}) + d\hat\sigma_{n+1} ]$ for exchangeable data  and $\mathcal{C}(\bx_{n+1}) =  [ \hat\mu(\x_{n+1}) + \hat\delta + \hat\rho \left\{ y_{n} - \hat\mu(\bx_{n}) \right\}  - d\hat\sigma_{n},  \hat\mu_\mathcal{A}(\x_{n+1}) + \hat\delta + \hat\rho \left\{ y_{n} - \hat\mu_\mathcal{A}(\bx_{n}) \right\}  + d\hat\sigma_{n} ]$ for time series data, where \(d\) is the \(k\)-th smallest value of \(\{R_i: i\in\mathcal I_2\}\), with $k=\big\lceil (|\mathcal I_2|+1)(1-\alpha)\big\rceil$. The resuting prediction interval is valid in finite samples for exchangeable data and asymptotically valid for time series data:

\begin{corollary} \label{thm:conformal_validity_split_adaptive}
	
(i) For exchangeable data, under Assumption~\ref{assump:exchangeable}, $\mathbb{P}\left\{y_{n+1}\in\mathcal{C}(\bx_{n+1})\right\}\geq 1-\alpha.$
		If, in addition, the fitted absolute residuals $|\widehat\eta_i|$, $i\in\{[n/2+1],\ldots,n+1\}$ have a continuous joint distribution all $y\in \mathbb{R}$, then it also holds that
		$\mathbb{P}\left\{y_{n+1}\in\mathcal{C}(\bx_{n+1})\right\}\leq 1-\alpha+1/(|\mathcal{I}_2|+1).$
(ii) For time series data, if Assumption \ref{assump:stationarity} hold and Assumptions \ref{assump:convergence}-\ref{assump:local} hold with $\boldsymbol \theta$ replaced by $(\boldsymbol \theta,  \boldsymbol \phi)$ and $e_i$ replaced by $\eta_i$, then
$
\mathbb{P}\left\{y_{n+1}\in\mathcal{C}(\bx_{n+1})\right\} \xrightarrow{p} 1 - \alpha.
$
\end{corollary}

In (i), the upper bound of coverage depends on the size of the second subsample rather than the full sample, reflecting the cost of sample splitting. In both (i) and (ii), because the parameters, including those for heteroskedasticity and serial correlation, are estimated from only half of the observations, the added estimation noise may lead to wider intervals. The proof follows the same arguments as Theorems~\ref{thm:conformal_validity_exchangeable} and~\ref{thm:conformal_validity_stationarity} and is omitted.

\subsection{Applications to model averaging}

Algorithms~\ref{algorithm:split_conformal_full_style} and~\ref{algorithm:split_conformal_adaptive} apply to all model averaging methods discussed in Subsection~4.3. They also allow the construction of prediction intervals with methods that are either too costly in their full-sample versions or that require resampling. We illustrate this with one example of each.

\noindent\textbf{Bayesian model averaging.} To apply Algorithms~\ref{algorithm:split_conformal_full_style} or~\ref{algorithm:split_conformal_adaptive} , split the index set ${\{1,\ldots,n\}}$  into a training set $\mathcal{I}_1$ and a calibration set $\mathcal{I}_2$, and denote the corresponding data by $\mathcal{D}_1$ and $\mathcal{D}_2$.  
For each candidate model $\mathcal{M}_m$, specify a prior $p(\boldsymbol{\theta}_m \mid \mathcal{M}_m)$ and a likelihood function.  
Using $\mathcal{D}_1$, compute the posterior mean $\hat{\boldsymbol{\theta}}_m$ from $p(\boldsymbol{\theta}_m \mid \mathcal{D}_1, \mathcal{M}_m)$ and the marginal likelihood:
\[
p(\mathcal{D}_1 \mid \mathcal{M}_m) = \int p(\mathcal{D}_1 \mid \boldsymbol{\theta}_m, \mathcal{M}_m) \, p(\boldsymbol{\theta}_m \mid \mathcal{M}_m) \, d\boldsymbol{\theta}_m.
\]  
The posterior model probability is then  
\[
\widehat{w}_m = \frac{p(\mathcal{D}_1 \mid \mathcal{M}_m) \cdot p(\mathcal{M}_m)}{\sum_{j=1}^M p(\mathcal{D}_1 \mid \mathcal{M}_j) \cdot p(\mathcal{M}_j)}.
\]
Next, apply $\{\hat{\boldsymbol{\theta}}_m, \widehat{w}_m\}_{m=1}^M$ to $\mathcal{I}_2$. For each $i \in \mathcal{I}_2$, the BMA prediction is  
\[
\hat{\mu}^{\text{BMA}}(\bx_i)= \sum_{m=1}^M \widehat{w}_m \, f^{(m)}(\bx_i; \hat{\boldsymbol{\theta}}_m),
\]  
and the conformity score is  
\[
R_{i} = \left| y_i - \hat{\mu}^{\text{BMA}}(\bx_i) \right|.
\]
Finally, compute the conformity score for any candidate value of $y_{n+1}$. The prediction interval is the set of candidate values whose conformity scores do not exceed the desired empirical quantile. If local adaptability is desired, an additional step can be included: obtain the heteroskedasticity or serial correlation coefficients at the posterior mean from $\mathcal{I}_1$, and then compute residuals for $\mathcal{I}_2$ using these estimates, as in Algorithm~\ref{algorithm:split_conformal_adaptive}.

The resulting prediction interval is valid even when all models are misspecified. This contrasts with conventional Bayesian credible sets, whose frequentist coverage depends on model specification. Although the interval is computed using a single parameter value (the posterior mean) and a single set of model weights, it does not understate uncertainty because it relies on conformity rather than the sampling distributions of the parameters and weights. This distinction is central to its validity.

\noindent\textbf{Bagging.} We consider a linear model setting for simplicity. In this context, bagging may involve generating multiple bootstrap resamples of the original prediction problem, applying a model selection rule to each resample, and averaging the resulting predictions. For example, \cite{inoue2008bagging} used this approach to forecast U.S. inflation.

Suppose $\bx_i \in \mathbb{R}^p$ and $y_i \in \mathbb{R}$, and we treat
\[
y_i = \bx_i^\top \boldsymbol{\beta} + e_i
\]
as the most general predictive model, with an unknown subset of $\bx_i$ potentially having predictive power. The model may be misspecified; for example, the true data-generating process could be nonlinear, or the model may omit important predictors.

To apply Algorithms~\ref{algorithm:split_conformal_full_style} and~\ref{algorithm:split_conformal_adaptive}, split the sample into two subsets of approximately equal size, as in the Bayesian model averaging case, with $\mathcal{I}_1$ as the training set and $\mathcal{I}_2$  the calibration set.  Fit bagged linear predictors on $\mathcal{I}_1$ possibly as follows. For $b = 1, \dots, B$: Draw a bootstrap sample of size $|\mathcal{I}_1|$ from $\mathcal{I}_1$ (sampling with replacement). Fit an OLS regression to the most general model on the bootstrap sample. Apply a method to determine which variables to include in this bootstrap sample. For example, one can perform individual $t$-tests on estimated coefficients against a fixed threshold and discard insignificant coefficients, as in \cite{inoue2008bagging}. Let $\mathbf{C}_b \in \mathbb{R}^{p \times p_b}$ be the resulting selection matrix such that $\mathbf{X} \mathbf{C}_b \in \mathbb{R}^{(n+1) \times p_b}$ contains only the selected columns of $\mathbf{X}$. Re-estimate the model using the selected regressors to obtain the final OLS estimate for this bootstrap sample, denoted $\hat{\boldsymbol{\beta}}^{(b)}$. The prediction at any $\mathbf{x} \in \mathbb{R}^p$ is then
\[
\hat{f}^{(b)}(\mathbf{x}) = \mathbf{x}^\top \mathbf{C}_b \, \hat{\boldsymbol{\beta}}^{(b)}.
\]
Repeat this for all $b = 1, \dots, B$, and average the predictions across bootstrap models to obtain
\[
\hat{f}(\mathbf{x}) = \frac{1}{B} \sum_{b=1}^B \hat{f}^{(b)}(\mathbf{x})
= \frac{1}{B} \sum_{b=1}^B  \mathbf{x}^\top \mathbf{C}_b \hat{\boldsymbol{\beta}}^{(b)}.
\]
All computations above use only the training set $\mathcal{I}_1$. Next, compute residuals on $\mathcal{I}_2$. For each $i \in \mathcal{I}_2$, the prediction, representing model averaging with equal weights, is given by
\[
\hat{\mu}(\bx_i) = \frac{1}{B} \sum_{b=1}^B  \mathbf{x}_i^\top \mathbf{C}_b \hat{\boldsymbol{\beta}}^{(b)}, \ \ (i \in \mathcal{I}_2),
\]
and the conformity score is
\[
R_{i} = \left| y_i - \hat{\mu}(\bx_i)\right|.
\]
Compute the conformity score for any candidate value of $y_{n+1}$. The prediction interval consists of candidate values whose conformity scores fall within the desired empirical quantile. The same remarks on local adaptability from the Bayesian model averaging case also apply here.

\section{Monte Carlo experiments}
We examine the coverage properties of the proposed algorithms for all model averaging methods in Subsection 4.3. The following methods are reported, each implemented as originally proposed without modification: equal weighting (Equal), the regression approach of \citeauthor{granger1984} (\citeyear{granger1984}) (Regression), Smoothed AIC (SAIC), Smoothed BIC (SBIC), Mallows’ model averaging (MMA), and Jackknife model averaging (JMA). We present results separately for the exchangeable and time-series cases. The coverage level of the prediction intervals is set at $90\%$, and each reported value is based on 500 replications.

\subsection{Experiment I: exchangeable data}
We consider the same DGP as in \citet{hansen2007}, with an extension to allow for heteroskedasticity. The DGP is an infinite-order regression model (so all candidate models are misspecified):
\[
y_i = \sum_{j=1}^{\infty} \beta_j x_{ji} + \sigma_i e_i,
\]
where $x_{1i} = 1$ is the intercept, the remaining $x_{ji}$ are i.i.d.\ $\mathcal{N}(0,1)$, and the error term $e_i \sim \mathcal{N}(0,1)$ is independent of $\bx_i$. The coefficients are
\[
\beta_j = c \sqrt{2\alpha}\, j^{-\alpha - 1/2};
\]
in the homoskedastic case ($\sigma_i = 1$), the population $R^2 = c^2/(1+c^2)$, controlled by $c$ and independent of $\alpha$. This corresponds exactly to the design in \citet{hansen2007}.  

We also consider a heteroskedastic extension in which the log standard deviation is
\[
\log(\sigma_i) = \sum_{j=1}^{\infty} \beta_j x_{ji},
\]
so that all variables potentially affect the variance. The same coefficients $\beta$ are used in both the mean and variance equations, which is not an issue here because the focus is on coverage properties of intervals rather than efficient estimation. This represents a more general setting with misspecification in both the mean and variance equations. 

The sample size is set to $n=150$, so the split-sample case has 75 training observations. 
The number of models $M$ follows \citet{hansen2007}, using the rule 
$M = \text{round}(3n^{1/3})$, which gives 16 models for this sample size. 
The $m$-th candidate model is  
\[
y_i = \sum_{j=1}^{m+1} \beta_j x_{ji} + \sigma_i e_i.
\]
For locally adaptive intervals, the log conditional variance is modeled as a linear function of all regressors in the largest candidate model. This reflects a conservative choice for local adaptive methods, since researchers may in practice use additional information to select covariates and obtain tighter intervals. We varied $\alpha = 0.5, 1.0,$ and $1.5$ as in \citet{hansen2007}$;$ larger $\alpha$ implies faster 
decay of the coefficients $\beta_j$ with $j$. 
We also considered different $c$ values, such that the population $R^2 = 0.2, 0.5,$ and $0.8$. 
To avoid repetition, results are reported only for $\alpha=1$ and $R^2=0.5$, since conclusions 
are similar across values.

\begin{figure}[htbp]
		 \caption{{Coverage for Exchangeable Data with Homogeneous Variance}}\label{fig:iid-ho-coverage}
		\centering\includegraphics[trim={0cm 0cm 0cm 1.4cm},clip, width=0.9\textwidth]{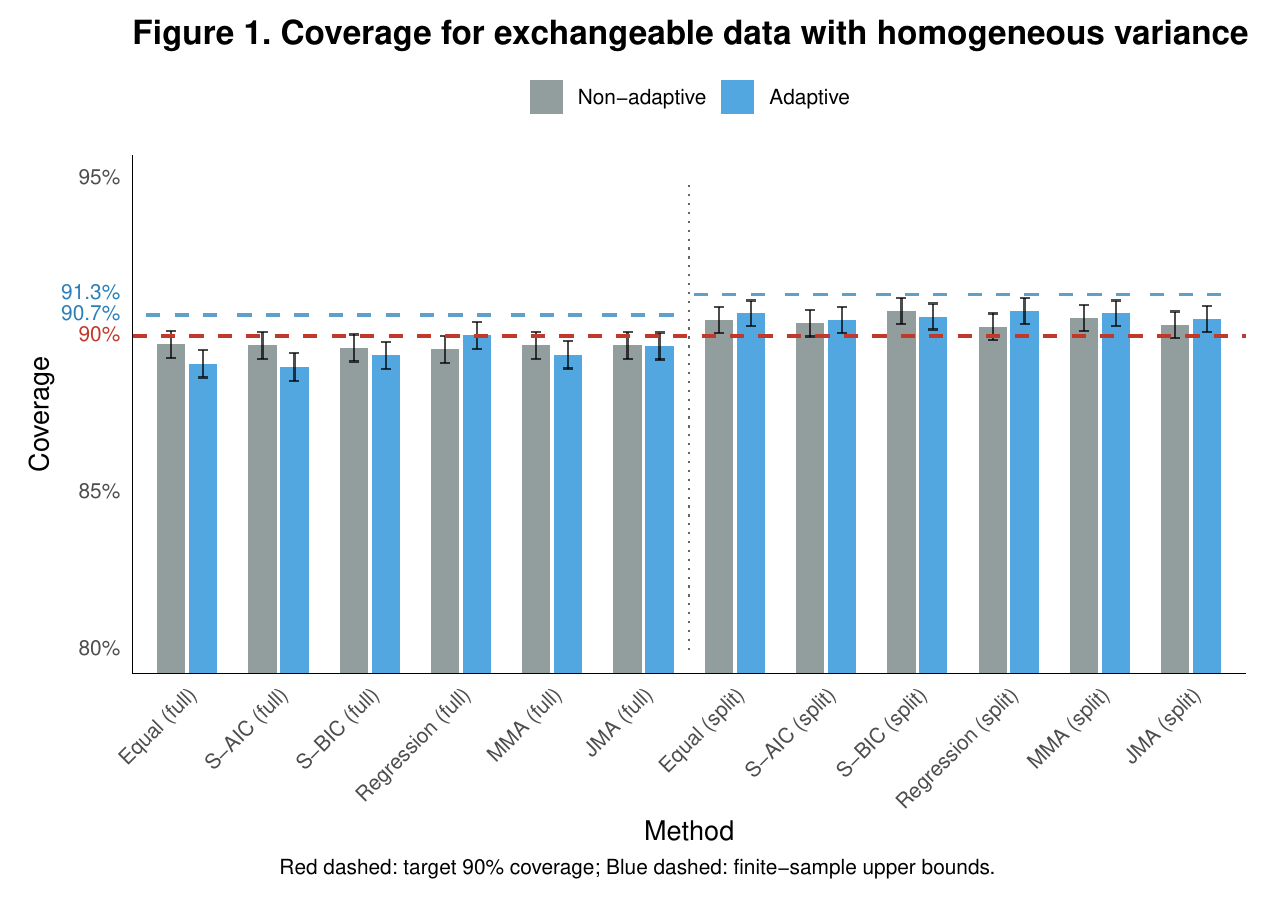} 

\end{figure}

\begin{figure}[htbp]
 \caption{{Interval Length for Exchangeable Data with Homogeneous Variance}}\label{fig:iid-ho-length}
	\centering	\includegraphics[trim={0cm 0cm 0cm 1.4cm},clip, width=0.9\textwidth]{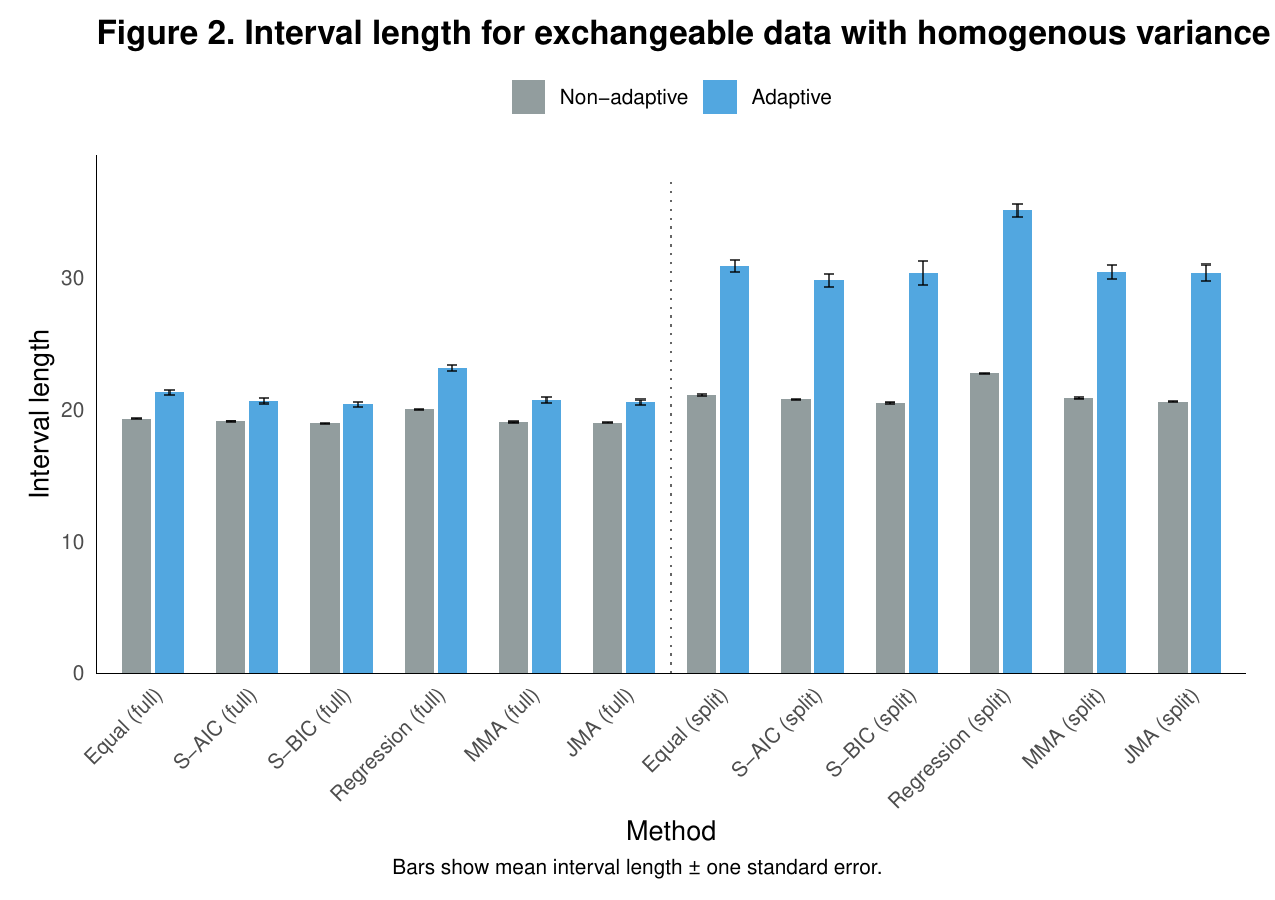} 
\end{figure}
Figures \ref{fig:iid-ho-coverage} and \ref{fig:iid-ho-length} report the coverage rates and interval lengths for the case with homogeneous variance. Figure \ref{fig:iid-ho-coverage} shows empirical coverage rates, with bars grouped into full-sample (left) and split-sample (right) implementations. Within each group, methods are further distinguished by whether they are adaptive to heteroskedasticity (blue) or non-adaptive (gray). Error bars represent one standard error. The red dashed line marks the 90\% target coverage, and the blue dashed lines indicate finite-sample upper bounds. Across all methods, coverage is close to or above the nominal level. Adaptive and non-adaptive versions perform similarly, showing that modeling heteroskedasticity does not compromise validity.

Figure \ref{fig:iid-ho-length} reports the corresponding average interval lengths in the same format. For non-adaptive methods, split-sample intervals are only slightly longer than full-sample intervals, suggesting a small efficiency loss from splitting. In contrast, adaptive methods produce noticeably wider intervals, reflecting the additional estimation variability introduced by modeling heteroskedasticity. The efficiency cost therefore stems less from sample splitting than from local adaptation. It is worth noting, however, that the effect of variance estimation will naturally depend on the application; the current design represents a conservative case, since all variables from the largest model are used to model the variance. Together, Figures 1 and 2 show that all methods maintain valid coverage, with efficiency tradeoffs depending on the chosen design.

\begin{figure}[htbp]
 \caption{{Coverage for Exchangeable Data with Hetergeneous Variance}}\label{fig:iid-he-coverage}
	\centering	\includegraphics[trim={0cm 0cm 0cm 1.4cm},clip, width=0.9\textwidth]{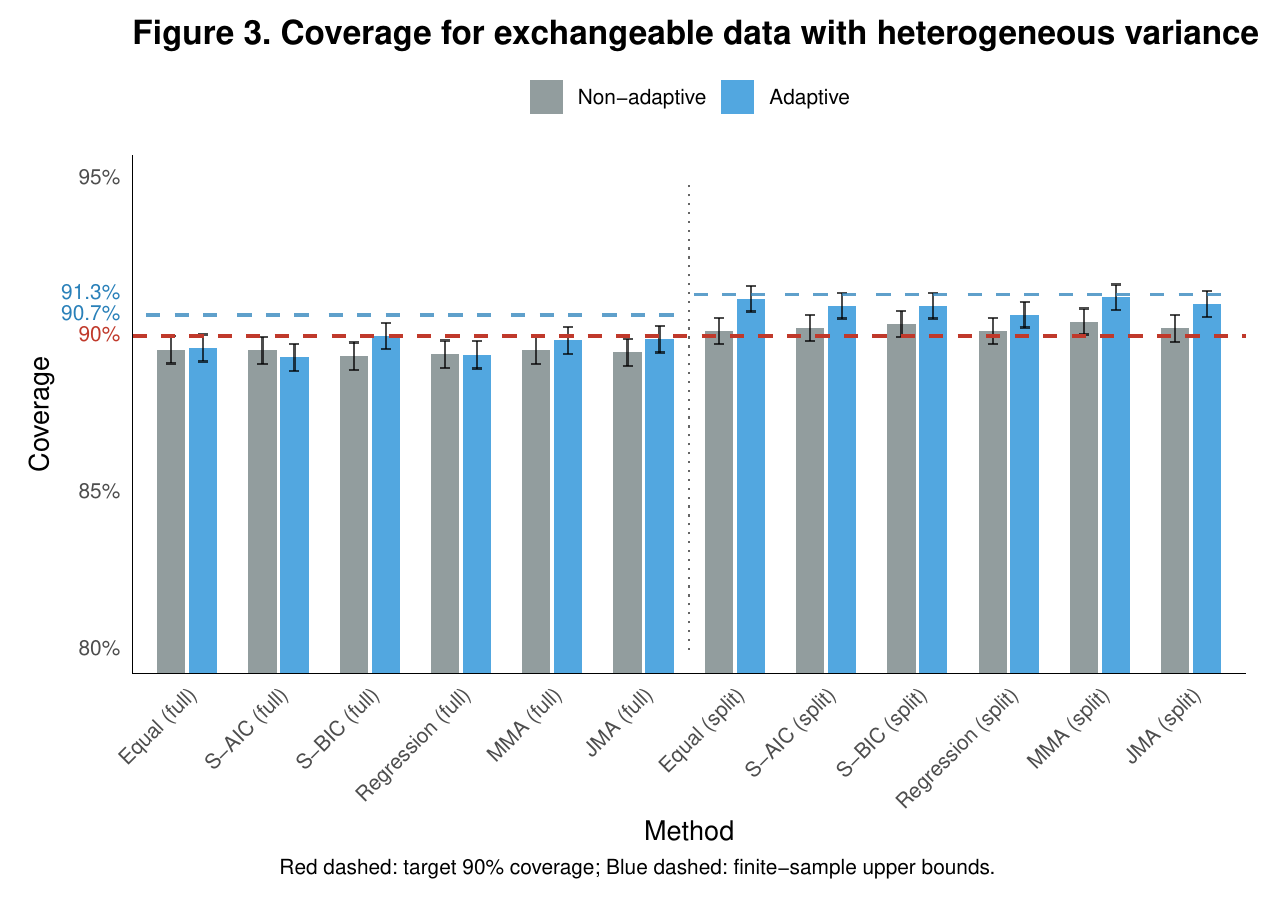} 
\end{figure}

\begin{figure}[htbp]
 \caption{{Interval Length for Exchangeable Data with Hetergeneous Variance}}\label{fig:iid-he-length}
	\centering	\includegraphics[trim={0cm 0cm 0cm 1.4cm},clip, width=0.9\textwidth]{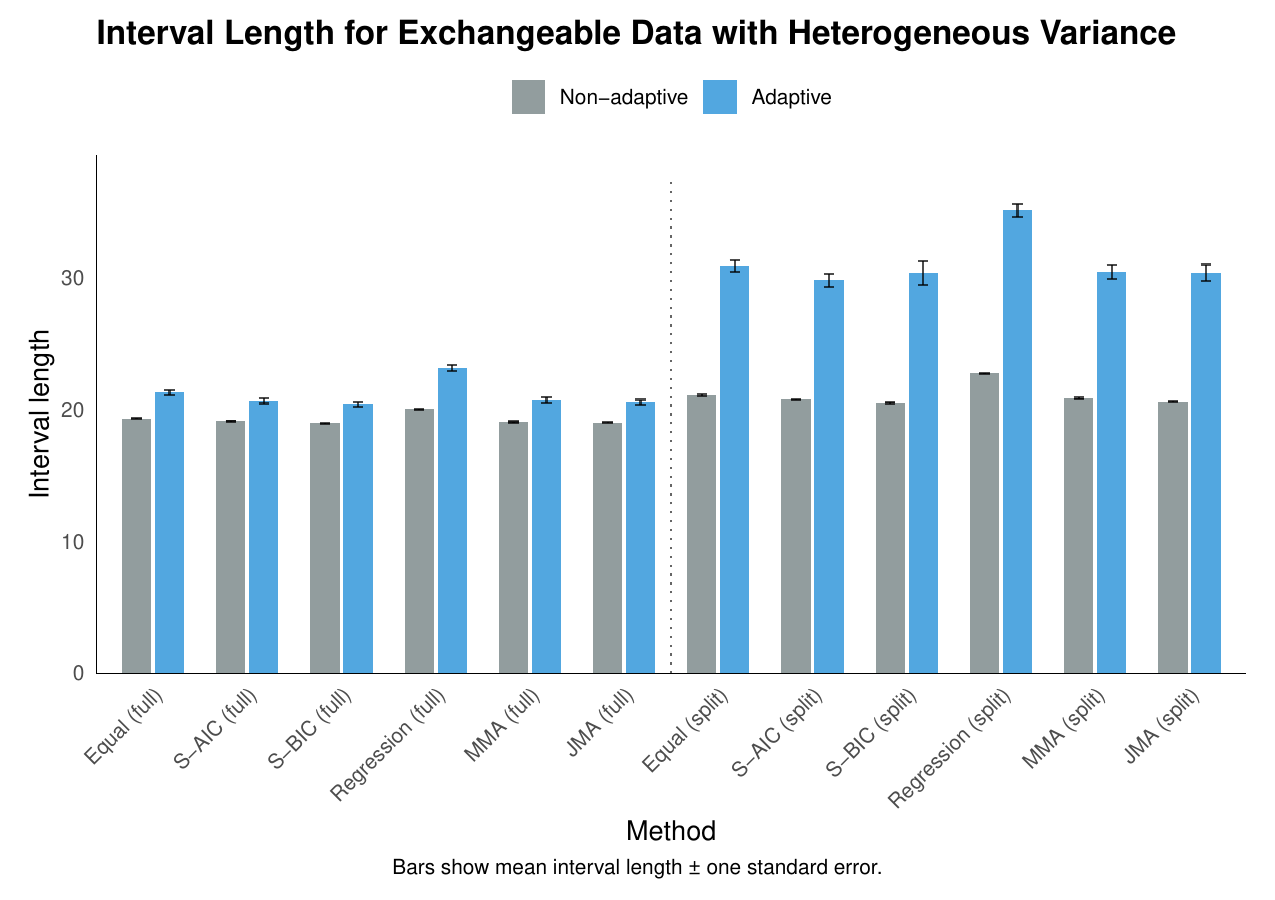} 
\end{figure}

Figures \ref{fig:iid-he-coverage} and \ref{fig:iid-he-length} present coverage and interval length when the data are generated with heterogeneous variance. Coverage remains close to or above the nominal level across all methods, much like in the homogeneous case, with little difference between adaptive and non-adaptive versions. Efficiency patterns are also similar: split-sample intervals are only slightly longer than full-sample ones for non-adaptive methods, while adaptive methods produce wider intervals. The main advantage of adaptive intervals is their ability to adjust to local variability, intervals should widen when variance is high and narrow when it is low, and we plan to include a closer examination of this pattern in the paper’s next version.

\subsection{Experiment II: time series data}
We design an experiment that mimics the empirical problem of predicting stock returns with many potential predictors, using parameter values calibrated to empirical estimates in Table~1 of \cite{boudoukh2008myth}. The data are generated from an infinite-order regression,  
\[
r_{i+1} = \delta + \sum_{j=1}^{\infty} \beta_j x_{j,i} + u_{i+1}, \quad 
x_{j,i+1} =  \rho_j x_{j,i} + v_{j,i+1}, \quad 
u_{i} = \sum_{j=1}^{\infty} \gamma_j v_{j,i}+e_{i},
\]
with $u_i \sim \text{i.i.d. } N(0,\sigma_u^2)$, $v_{j,i} \sim \text{i.i.d. } N(0,\sigma_v^2)$ independent across$ j$, and $e_i \sim \text{i.i.d. } N(0,\sigma_e^2)$. The coefficients  
\[
\beta_j = c \sqrt{2\alpha}\, j^{-\alpha - 1/2}
\]
decay only gradually, so that many predictors contribute to predictability. Each predictor follows an AR(1) process, a common assumption in the return-prediction literature. The errors $u_i$ and $v_{j,i}$ are set up to be negatively correlated, which is known to bias estimates of $\beta_j$, a feature emphasized in earlier empirical work, including Table~1 of \cite{boudoukh2008myth}.

\begin{figure}[htbp]
 \caption{{Coverage for Time-Series Data with Homogeneous Variance (n=100)}}\label{fig:iid-ts-coverage}
	\centering	\includegraphics[trim={0cm 0cm 0cm 1.4cm},clip, width=0.9\textwidth]{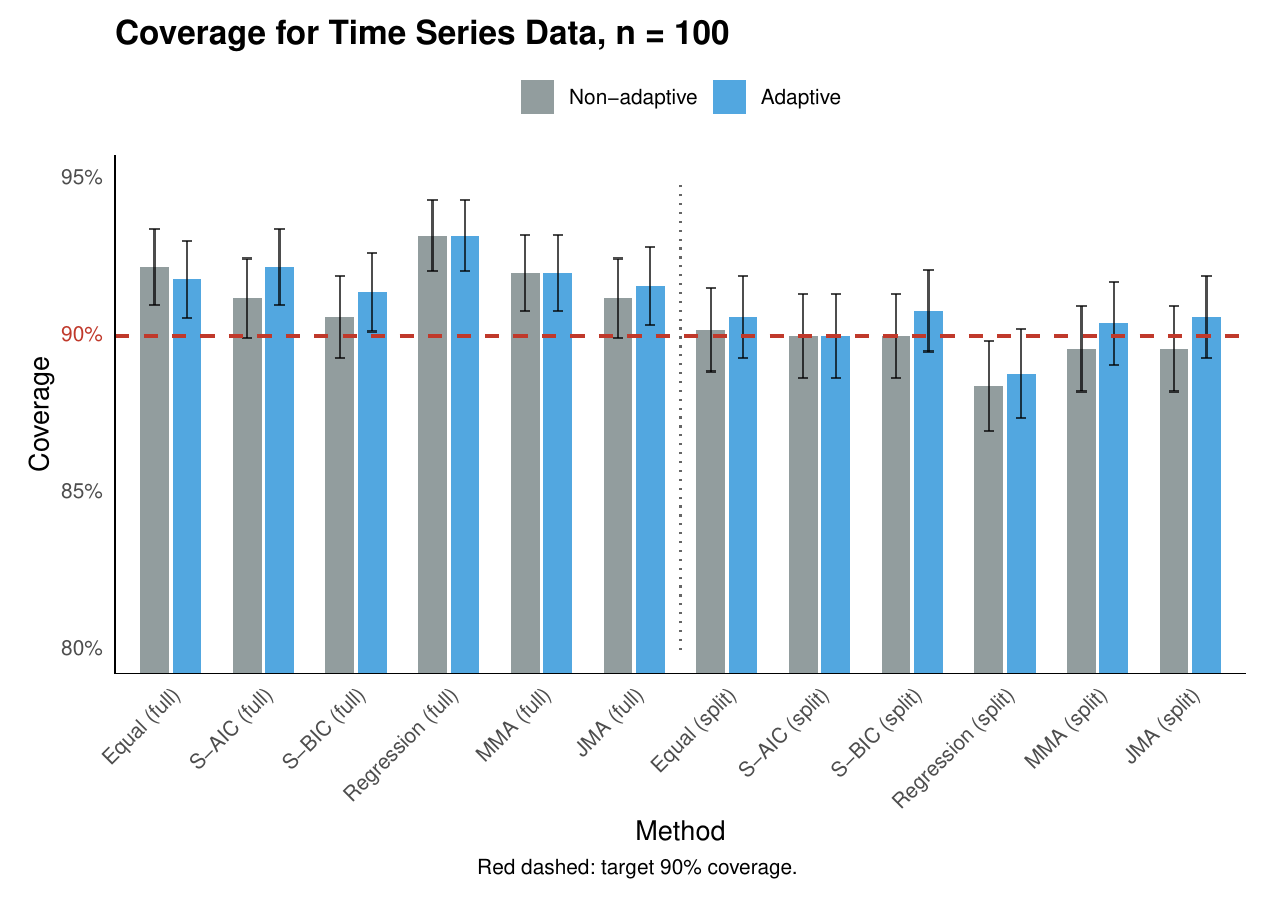} 
\end{figure}

We set $\delta = 0$ (without loss of generality), take $\rho_j = 0.953$ for all $j$ (so all regressors are persistent), and set $\sigma_{u} = 0.202$ and $\sigma_{v} = 0.154$. The correlation $\text{corr}(v_{j,i}, u_i) = -0.712$ implies a projection coefficient $\text{cov}(u_i, v_{j,i})/\text{var}(v_{j,i}) \approx -0.933$; $\gamma_j$ is drawn uniformly from $[-0.933,0]$;  and $\sigma_e = \sqrt{\sigma_u^2 - \sigma_v^2 \sum_{j=1}^{\infty} \gamma_j^2}$. This specification coincides with the DGP in \citet{boudoukh2008myth} when there is only a single predictor and $\gamma=\gamma_j=-0.933$. Next, we set $\alpha=1$ as in the exchangeable simulation case and choose $c$ so that the $R^2$ of the return prediction equation, $R^2=1-\text{var}(u_i)/\text{var}(r_i)$, equals $0.2$, consistent with weak empirical predictability. Results for $R^2=0.5$ and $0.8$ are similar and are omitted.

\begin{figure}[htbp]
 \caption{{Interval Length for Time-Series Data with Homogeneous Variance (n=100)}}\label{fig:ts-he-length}
	\centering	\includegraphics[trim={0cm 0cm 0cm 1.4cm},clip, width=0.9\textwidth]{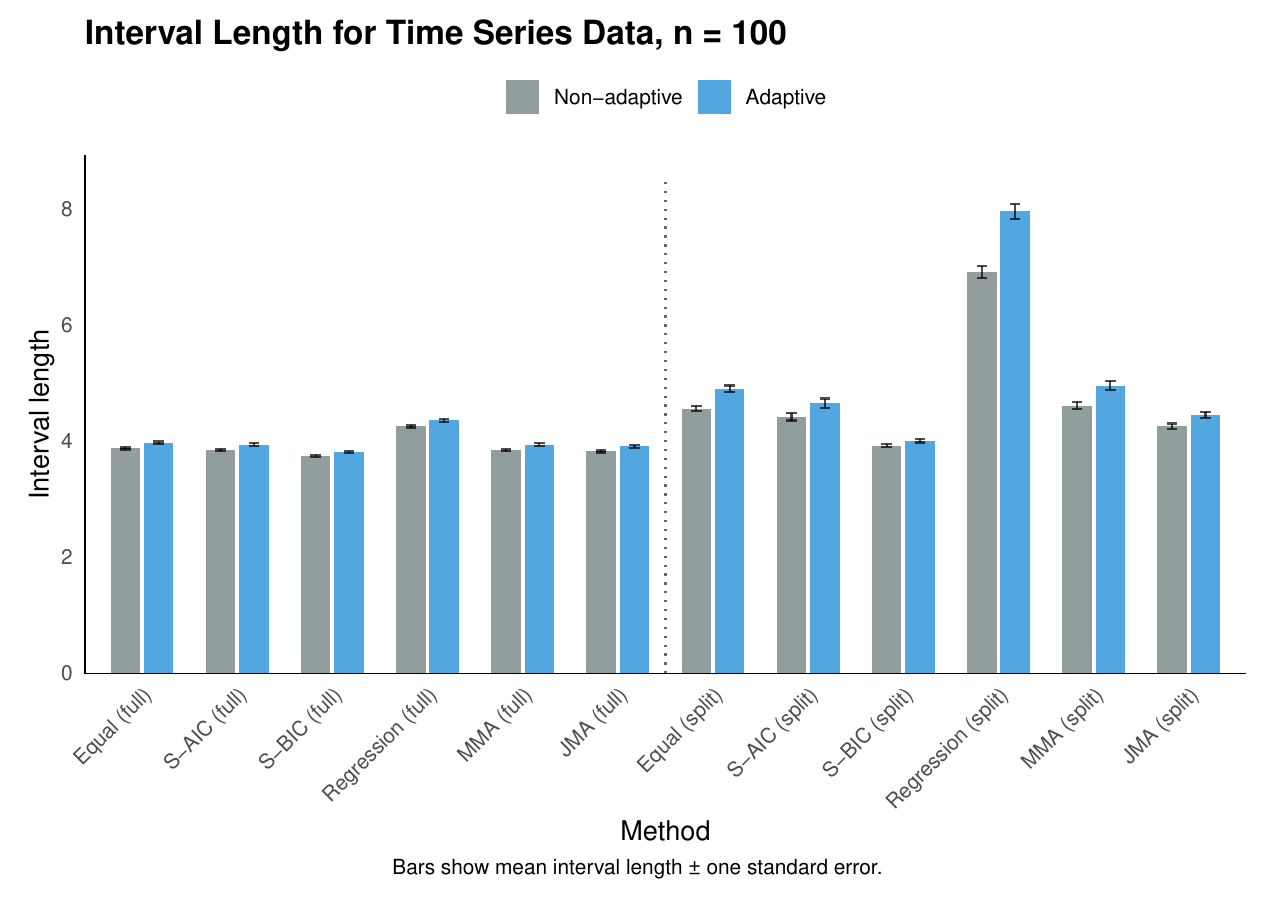} 
\end{figure}

We consider two sample sizes, $n = 100$ and $n = 200$, with split-sample training sets of 50 and 100 observations, respectively. The number of models follows the rule in \citet{hansen2007}, $M = \text{round}(3n^{1/3})$. The $m$-th candidate model is
\[
r_{i+1} = \delta + \sum_{j=1}^{m+1} \beta_j x_{j,i} + u_{i+1}.
\]
Figures 5 and 6 report coverage and interval length for the time-series design ($n=100$). Coverage remains at or above the 90\% target across all methods, with little distinction between adaptive and non-adaptive versions. Interval length shows the expected small loss from sample splitting, but the gap between adaptive and non-adaptive intervals is much smaller compared with the exchangeable case. This reflects the design: the AR(1)–GARCH(1,1) model requires far fewer parameters to estimate than the high-dimensional specification in the exchangeable case, so local adaptation adds less variability. Overall, adequate coverage is observed even for such a small sample size, splitting costs are small, and the efficiency penalty from adaptation is milder in the time-series setting. Among the methods, however, the regression-based approaches yield noticeably wider intervals and somewhat lower coverage, a pattern that warrants further investigation.

\begin{figure}[htbp]
 \caption{{Coverage for Time-Series Data with Homogeneous Variance (n=200)}}\label{fig:iid-ts-coverage-2}
	\centering	\includegraphics[trim={0cm 0cm 0cm 1.4cm},clip, width=0.9\textwidth]{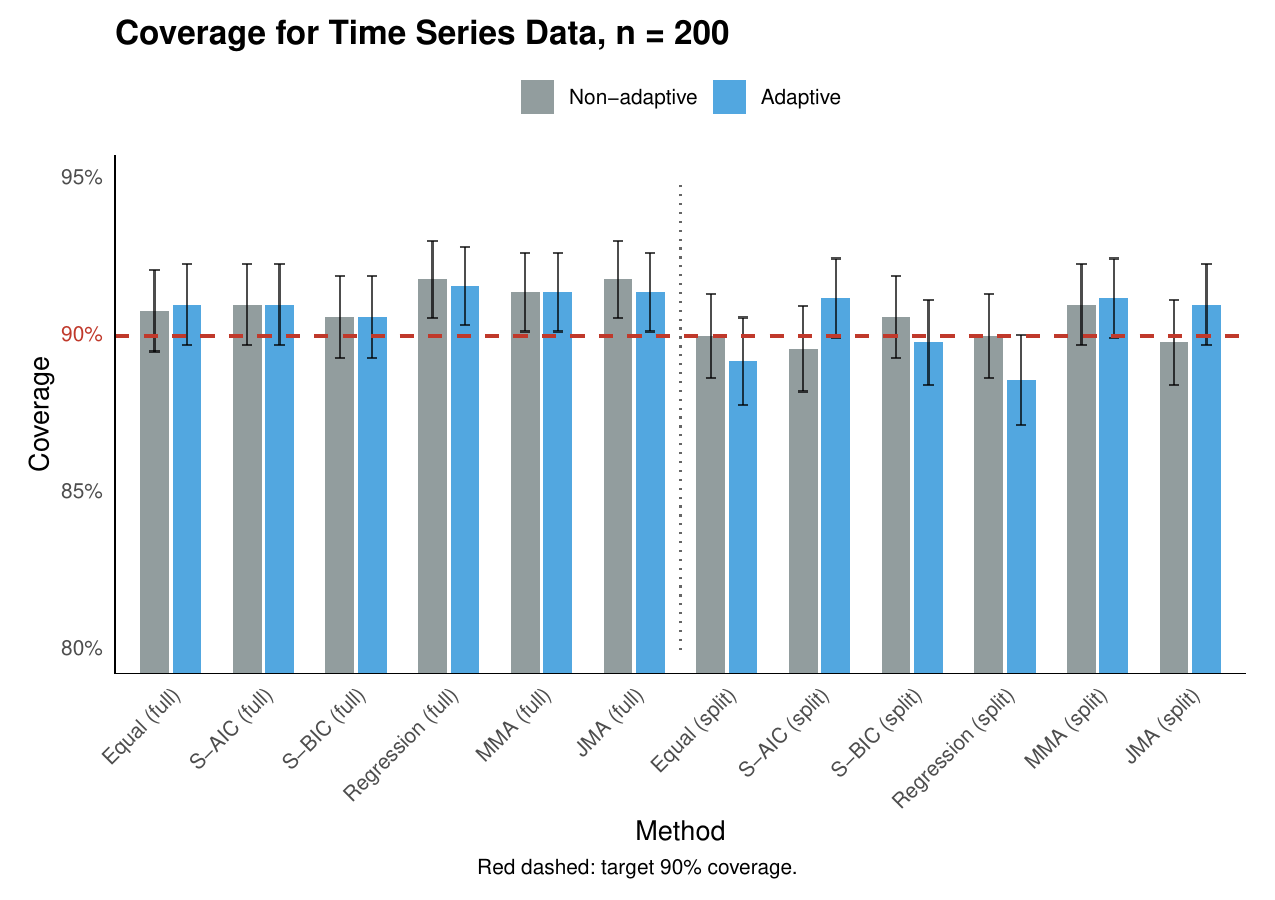} 
\end{figure}

\begin{figure}[htbp]
 \caption{{Interval Length for Time-Series Data with Homogeneous Variance (n=200)}}\label{fig:ts-he-length-2}
	\centering	\includegraphics[trim={0cm 0cm 0cm 1.4cm},clip, width=0.9\textwidth]{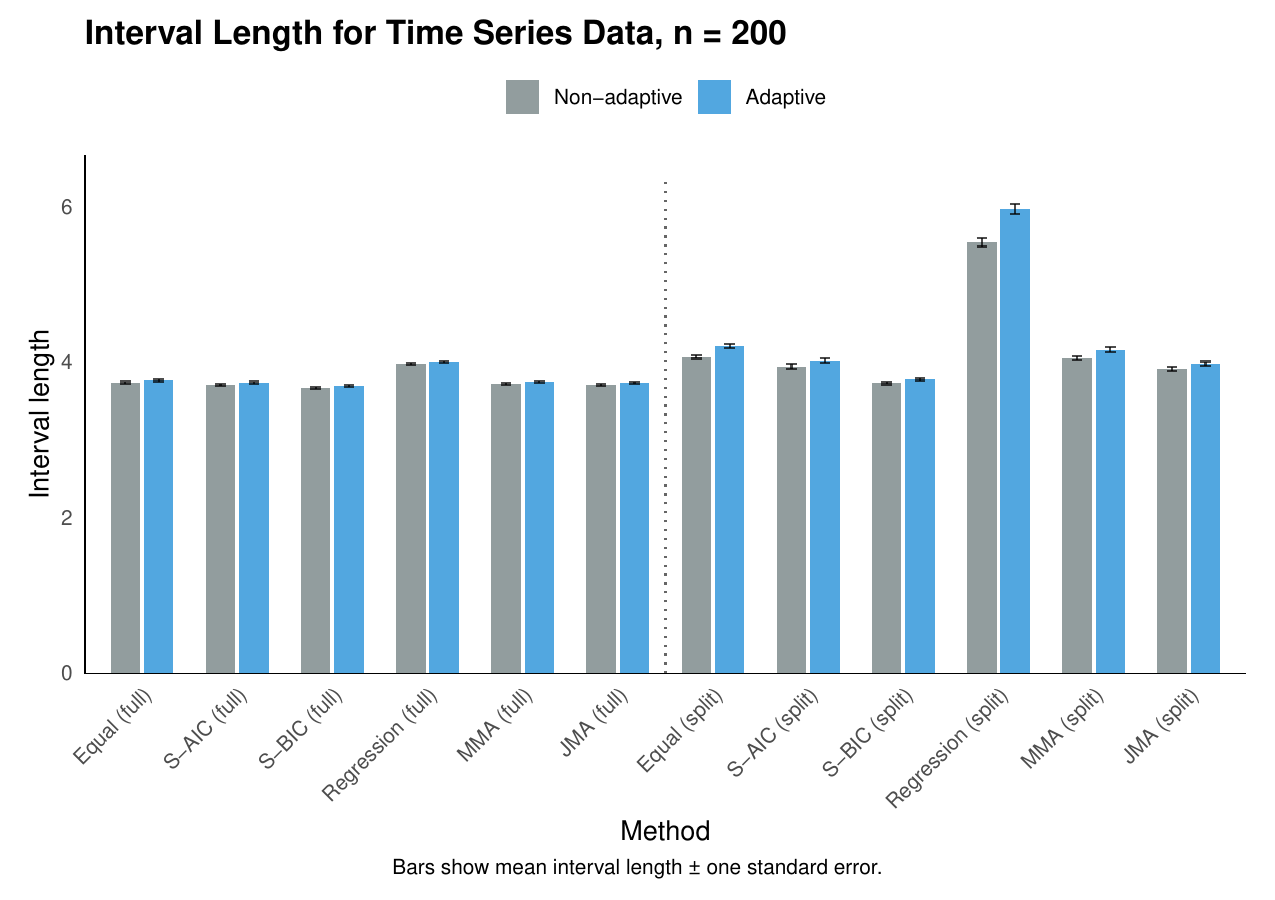} 
\end{figure}

Figures \ref{fig:iid-ts-coverage-2} and \ref{fig:ts-he-length-2} present results for the time-series design with $n=200$. Coverage remains close to or above the 90\% target and is more stable than in the $n=100$ case. Interval lengths again show a slight cost from sample splitting and a modest gap between adaptive and non-adaptive methods. Regression-based methods still yield wider intervals and lower coverage, though the gap is smaller than in the $n=100$ case. Results for $n=400$ (omitted) show further improvements along the same lines.

Taken together, these results show that the methods achieve adequate coverage in a challenging setting with small samples, model misspecification, biased parameter estimates, persistent regressors, and many predictors. Further work on their performance under stochastic volatility is currently in progress.

\section{Empirical Applications}
Since the proposed algorithms are intended for both cross-sectional and time-series applications, we consider two empirical applications in these respective settings to evaluate their performance. We do not focus on efficient prediction, but on coverage properties in complex situations with limited sample sizes and either many candidate models or many potential predictors, as often encountered in empirical work.

\subsection{Real estate appraisal}

Housing is the largest asset for most households. Many predictors of housing values, such as location, proximity to facilities, house age, and transaction date, have been proposed, but their effects are often uncertain and vary across datasets. Model averaging offers a natural way to address this issue and potentially improve predictive accuracy. Yet accurate prediction is only part of the picture: a point estimate of 1 million with a 95\% interval of [0.9, 1.1] has very different implications than [0.5, 1.5]. In this section, we illustrate how our methods can provide both point estimates and prediction intervals to capture this uncertainty.
\renewcommand{\arraystretch}{1.0} 
\begin{sidewaystable}[!htbp]
\centering
\caption{Real estate appraisal with 95\% and 90\% prediction intervals}\label{tab:emp-cs}
\resizebox{\textheight}{!}{
\begin{tabular}{@{}l cc ccc ccc ccc ccc@{}}
\toprule
 & & & \multicolumn{6}{c}{95\%} & \multicolumn{6}{c}{90\%} \\
\cmidrule(lr){4-9} \cmidrule(lr){10-15}
\textbf{Method} & \textbf{RMSPE} & \textbf{Hit} &
\multicolumn{3}{c}{Non-adaptive} & \multicolumn{3}{c}{Adaptive} &
\multicolumn{3}{c}{Non-adaptive} & \multicolumn{3}{c}{Adaptive} \\
 & & & Coverage & Ave.Len & Sd.Len & Coverage & Ave.Len & Sd.Len
 & Coverage & Ave.Len & Sd.Len & Coverage & Ave.Len & Sd.Len \\
\midrule
Equal (full)  & 9.28 & 0.64 & 0.95 & 32.67 & 0.202 & 0.95 & 33.99 & 8.790 & 0.90 & 25.90 & 0.336 & 0.90 & 26.43 & 6.773 \\
Equal (split) &      &      & 0.96 & 33.87 & 0.324 & 0.94 & 32.66 & 8.513 & 0.90 & 25.34 & 0.135 & 0.90 & 25.85 & 6.747 \\
\addlinespace
Reg (full)    & 8.94 & 0.70 & 0.95 & 31.57 & 0.446 & 0.95 & 34.28 & 10.188 & 0.90 & 23.65 & 0.316 & 0.90 & 25.30 & 7.567 \\
Reg (split)   &      &      & 0.95 & 32.92 & 0.428 & 0.95 & 34.19 & 6.860  & 0.91 & 23.88 & 0.190 & 0.90 & 22.25 & 4.368 \\
\addlinespace
SAIC (full)   & 8.93 & 0.70 & 0.95 & 31.54 & 0.358 & 0.95 & 33.58 & 9.532  & 0.90 & 23.64 & 0.256 & 0.90 & 25.09 & 7.231 \\
SAIC (split)  &      &      & 0.96 & 33.47 & 0.517 & 0.95 & 32.18 & 6.664  & 0.91 & 24.45 & 0.284 & 0.89 & 22.62 & 4.572 \\
\addlinespace
SBIC (full)   & 8.93 & 0.70 & 0.95 & 31.72 & 0.369 & 0.95 & 33.38 & 8.960  & 0.90 & 23.61 & 0.269 & 0.90 & 24.63 & 6.832 \\
SBIC (split)  &      &      & 0.96 & 34.22 & 0.608 & 0.95 & 32.12 & 7.690  & 0.91 & 24.43 & 0.091 & 0.90 & 23.40 & 5.502 \\
\addlinespace
MMA (full)    & 8.93 & 0.70 & 0.95 & 31.46 & 0.478 & 0.95 & 33.74 & 8.443  & 0.90 & 23.76 & 0.249 & 0.90 & 24.33 & 6.068 \\
MMA (split)   &      &      & 0.96 & 32.85 & 0.416 & 0.95 & 33.95 & 8.117  & 0.91 & 24.63 & 0.120 & 0.90 & 24.06 & 5.786 \\
\addlinespace
JMA (full)    & 8.93 & 0.70 & 0.95 & 31.37 & 0.511 & 0.95 & 34.07 & 8.279  & 0.90 & 23.80 & 0.264 & 0.90 & 24.34 & 5.941 \\
JMA (split)   &      &      & 0.95 & 32.64 & 0.347 & 0.95 & 33.71 & 7.622  & 0.91 & 24.38 & 0.121 & 0.90 & 24.23 & 5.419 \\
\addlinespace
LG (full)     & 8.94 & 0.70 & 0.95 & 31.57 & 0.447 & 0.95 & 34.29 & 10.186 & 0.90 & 23.65 & 0.316 & 0.91 & 25.31 & 7.586 \\
LG (split)    &      &      & 0.95 & 32.92 & 0.428 & 0.95 & 34.20 & 6.877  & 0.91 & 23.88 & 0.190 & 0.90 & 22.25 & 4.371 \\
\bottomrule
\end{tabular}
}
\caption*{\footnotesize \textit{Notes}: RMSPE is root mean squared prediction error. Coverage is the empirical coverage probability of the prediction intervals. Ave.Len and Sd.Len are the average and standard deviation of interval lengths. “Full” and “split” indicate whether the full training sample is used or split into halves (for parameter and weight estimation). ``Adaptive'' intervals adjust for serial dependence and heteroskedasticity; ``non-adaptive'' intervals do not. Equal = equal weighting, Reg = the regression approach, SAIC = smoothed AIC weights, SBIC = smoothed BIC weights, MMA = Mallows model averaging, JMA = jackknife model averaging, LG= the largest model.}
\end{sidewaystable}

We consider housing prices in Taipei using the dataset of \citet{yeh2018}, available through the UC Irvine Machine Learning Repository. The dataset contains 414 observations, with the outcome variable defined as the residential housing price per unit area. Prices were collected from the Ministry of the Interior’s public database for the period June 2012 to May 2013, covering two districts in Taipei City and two districts in New Taipei City. Six appraisal predictors are included: distance to the nearest MRT station, number of convenience stores within walking distance, house age, transaction date, and geographic coordinates (latitude and longitude). See \citet{yeh2018} for details on the construction of these variables. For prediction, we use a leave-one-out procedure: each property is in turn treated as the test observation, with the remaining 413 serving as the training sample. We also experiment with randomly selecting 100 test observations, and the results are similar.

We average over alternative candidate models that differ in their choice of covariates. Specifically, we estimate the following regressions
$$
y_{i}=\boldsymbol{\beta}_m^\prime\bx_i^{(m)} +e_i,
$$
where $y_i$ denotes the transaction price per unit area for the $i$-th property, and $\bx_i^{(m)}$ is a vector of potential predictors of its price, which always contains the intercept but differs in observable characteristics. We consider all possible combinations of six appraisal factors, leading to $63=2^6-1$ candidate models. Predictions from these models are combined using six model averaging methods: equal weighting, regression-based weighting, SAIC, SBIC, MMA, and JMA. For comparison, we also report results from the largest model containing all predictors, a common practice in the price-prediction literature. The prediction interval for the largest model is computed as in the general case by setting the weight on the largest model to 1 and all others to 0.

We evaluate the methods through 95\% and 90\% coverage and average and standard error of the interval length. We also report root mean squared prediction error (RMSPE) and 20\% hit rate of the point prediction, both standard criteria in house appraisal studies, and defined as
\[\text{RMSPE}= \sqrt{ \frac{1}{N} \sum_{i=1}^N \mathbf{1}\!\left(\hat{y}_i - y_i \right)^2}, \quad\text{and}\quad
\text{Hit} = \frac{1}{N} \sum_{i=1}^N \mathbf{1}\!\left( \frac{| \hat{y}_i - y_i |}{y_i} \leq \tau \right),
\]
where $\tau=0.2$ is the tolerance threshold. Reported statistics are averages over all 414 evaluation points and are reported in Table~\ref{tab:emp-cs}.  

We find that the RMSPE of point predictions is broadly similar across methods, with the notable exception of equal-weight averaging. Equal weighting underperforms because most predictors are informative and larger models generally produce more accurate forecasts; giving the same weight to both small and large models leads to suboptimal results. The other methods perform comparably to the largest model, in some cases achieving slight improvements. This is encouraging because it shows that model averaging tilts the weights toward richer models even with 63 candidates to sift through.

All methods achieve coverage close to the nominal levels, with deviations at most 1\%. Sample splitting only slightly increases interval length, consistent with our simulations. With adaptive conformal methods, intervals are only mildly wider on average, suggesting a small efficiency loss. Meanwhile, the variation in interval lengths is strikingly different: without adaptation, lengths are essentially constant across predictions, whereas with adaptation their standard deviation is often comparable to the RMSPE. This is not a defect. By construction, adaptive intervals are meant to vary with the conditional variance of the residual, and here some covariates help explain that variance. In particular, a simple plot of interval length against property value shows narrower intervals for lower-priced properties and wider intervals otherwise, a desirable feature from a valuation perspective.

To summarize, for this cross-sectional application, coverage is accurate when averaging over a large number of models that include both nested and non-nested specifications. The resulting weights are informative, tilting toward richer and more predictive specifications. Split-sample procedures achieve efficiency comparable to full-sample procedures at much lower computational cost. Finally, adaptive intervals yield shorter intervals where variance is low and longer where it is high, while preserving nominal coverage and only mildly increasing average length. We next consider a time series application.

			
			
			
			
			
			
			

\subsection{Equity premium prediction}
Prediction of the equity premium (the difference between equity returns and the risk-free rate) has long been a central topic in finance. Many predictors have been proposed, each showing predictive power in some contexts, but none consistently dominates. A natural consideration is to apply model averaging across predictors, while quantifying prediction uncertainty. To our knowledge, however, no studies have offered frequentist prediction intervals in this setting. Our analysis provides such results.

We use the dataset and model specifications from \citet{goyal2024}, which reexamined 29 predictors proposed in recent papers (listed their Table~1), of which 20 are available at the monthly frequency. Our analysis includes all 20 monthly predictors. Their study, as is common in this literature, focused on bivariate regressions that include a single predictor at a time:  
\[
r_{t+1} = \alpha_i + \beta_i x_{i,t} + u_{i,t+1},
\]  
where $r_{t+1}$ denotes the equity premium (the value-weighted S\&P 500 return including dividends minus the one-month T-bill rate), and $x_{i,t}$ is predictor $i$, as defined in Table~2 of \citet{goyal2024}. We adopt the same practice, obtaining 20 candidate models for averaging. The sample period follows their study but begins in January 1996 to ensure all predictors are available, producing a balanced panel of 312 monthly observations through December 2021, with forecasts from all candidate models at each point.

\renewcommand{\arraystretch}{1.0} 
\begin{sidewaystable}[!htbp]
\centering
\caption{Equity premium prediction with 95\% and 90\% prediction intervals}\label{tab:emp-ts}
\resizebox{\textheight}{!}{
\begin{tabular}{l c ccc ccc ccc ccc}
\toprule
 & \multirow{2}{*}{MSPE}
 & \multicolumn{6}{c}{95\% coverage}
 & \multicolumn{6}{c}{90\% coverage} \\
\cmidrule(lr){3-8}\cmidrule(lr){9-14}
 &  & \multicolumn{3}{c}{Non-adaptive} & \multicolumn{3}{c}{Adaptive}
    & \multicolumn{3}{c}{Non-adaptive} & \multicolumn{3}{c}{Adaptive} \\
 &  & Coverage & Ave.Len & Sd.Len & Coverage & Ave.Len & Sd.Len
    & Coverage & Ave.Len & Sd.Len & Coverage & Ave.Len & Sd.Len \\
\midrule
Equal (full)  & 0.0016 & 0.95 & 0.18 & 0.004 & 0.95 & 0.18 & 0.004 & 0.91 & 0.15 & 0.003 & 0.91 & 0.15 & 0.003 \\
Equal (split) &        & 0.94 & 0.17 & 0.014 & 0.94 & 0.18 & 0.010 & 0.90 & 0.15 & 0.017 & 0.91 & 0.15 & 0.015 \\
\\
Reg (full)    & 0.0020 & 0.96 & 0.18 & 0.015 & 0.94 & 0.18 & 0.017 & 0.91 & 0.14 & 0.015 & 0.90 & 0.15 & 0.014 \\
Reg (split)   &        & 0.96 & 0.19 & 0.021 & 0.95 & 0.20 & 0.029 & 0.91 & 0.15 & 0.021 & 0.91 & 0.15 & 0.024 \\
\\
SAIC (full)   & 0.0020 & 0.94 & 0.17 & 0.016 & 0.94 & 0.18 & 0.017 & 0.91 & 0.14 & 0.015 & 0.91 & 0.15 & 0.014 \\
SAIC (split)  &        & 0.95 & 0.18 & 0.018 & 0.95 & 0.19 & 0.017 & 0.91 & 0.15 & 0.016 & 0.91 & 0.15 & 0.016 \\
\\
SBIC (full)   & 0.0020 & 0.94 & 0.17 & 0.016 & 0.94 & 0.18 & 0.017 & 0.91 & 0.14 & 0.015 & 0.91 & 0.15 & 0.014 \\
SBIC (split)  &        & 0.95 & 0.18 & 0.018 & 0.95 & 0.19 & 0.017 & 0.91 & 0.15 & 0.016 & 0.91 & 0.15 & 0.016 \\
\\
MMA (full)    & 0.0020 & 0.96 & 0.18 & 0.015 & 0.96 & 0.18 & 0.016 & 0.91 & 0.14 & 0.015 & 0.91 & 0.15 & 0.014 \\
MMA (split)   &        & 0.96 & 0.19 & 0.021 & 0.95 & 0.19 & 0.023 & 0.91 & 0.15 & 0.016 & 0.91 & 0.15 & 0.020 \\
\\
JMA (full)    & 0.0019 & 0.95 & 0.18 & 0.014 & 0.94 & 0.18 & 0.014 & 0.91 & 0.14 & 0.015 & 0.91 & 0.15 & 0.013 \\
JMA (split)   &        & 0.96 & 0.19 & 0.018 & 0.95 & 0.19 & 0.020 & 0.90 & 0.15 & 0.020 & 0.90 & 0.15 & 0.020 \\
\\
RW (full)     & 0.0016 & 0.95 & 0.18 & 0.003 & 0.95 & 0.18 & 0.002 & 0.93 & 0.15 & 0.004 & 0.91 & 0.15 & 0.003 \\

RW (split)    &        & 0.95 & 0.17 & 0.014 & 0.94 & 0.18 & 0.010 & 0.90 & 0.15 & 0.017 & 0.90 & 0.15 & 0.015 \\
\bottomrule
\end{tabular}
}
\caption*{\footnotesize \textit{Notes}: MSPE is mean squared prediction error. Coverage is the empirical coverage probability of the prediction intervals. Ave.Len and Sd.Len are the average and standard deviation of interval lengths. “Full” and “split” indicate whether the full training sample is used or split into halves (for parameter and weight estimation). ``Adaptive'' intervals adjust for serial dependence and heteroskedasticity; ``non-adaptive'' intervals do not. Equal = equal weighting, Reg = the regression approach, SAIC = smoothed AIC weights, SBIC = smoothed BIC weights, MMA = Mallows model averaging, JMA = jackknife model averaging, RW = random walk. Note that SAIC=SBIC because all models contain the same number of regressors.}

\end{sidewaystable}

We use a rolling-window design common in this literature. A fixed sample of 212 observations is moved forward one period at a time, with the subsequent observation serving as the prediction target, which produces 100 predictions in total. For methods involving sample splitting, half of the 212 observations are used for estimation. For all methods, we report the mean squared prediction error (MSPE) of the point prediction, the 95\% and 90\% coverage rate, and the mean and standard deviation of interval lengths. All statistics are averaged over the 100 predictions.

Table~\ref{tab:emp-ts} shows that equal weighting achieves the lowest MSPE, and the remaining MSPEs are similar. This general pattern suggests that the uncertainty associated with estimating 19 model weights outweighs the potential gains from unequal weighting, even if such weighting would be optimal in principle. 

More central to our analysis are the coverage properties of the prediction intervals. All methods deliver coverage close to the nominal level: between 0.94 and 0.96 at the 95\% level, and between 0.90 and 0.91 at the 90\% level. Average interval lengths are also similar across methods, roughly 0.18–0.19 for 95\% intervals and 0.14–0.15 for 90\% intervals.

A sharper difference appears in the variability of interval lengths. The split-sample versions consistently show larger standard deviations, in some cases much higher than those of the full-sample versions. This arises from dividing the estimation sample, which increases variation even though average lengths remain similar. Adaptive intervals are slightly more variable than their non-adaptive counterparts, suggesting limited gains or losses from allowing adaptivity, since the rolling-window scheme already provides some adjustment. These findings are consistent with the simulation evidence documented in the previous section.

Taken together, the results suggest that model averaging yields prediction intervals with reliable coverage in this time-series setting and broadly comparable efficiency across methods. Equal weighting performs favorably relative to more elaborate schemes, though in settings with fewer models or larger samples, unequal weighting may perform better. Our methods can readily be applied to such settings without modification.

\section{Conclusion}

We have proposed prediction intervals for general model averaging using conformal inference. These intervals cover out-of-sample realizations of the outcome variable with a chosen probability, thereby quantifying predictive uncertainty beyond point prediction. A point prediction without a measure of uncertainty offers limited guidance for empirical analysis and decision making. Our framework allows for general model misspecification and applies to averaging across multiple candidate models—nested, disjoint, overlapping, or any combination—with weights that may depend on the estimation sample. We established coverage guarantees under two sets of assumptions: exact finite-sample validity under exchangeability, relevant for cross-sectional data, and asymptotic validity under stationarity, relevant for time series. We described a benchmark algorithm, along with locally adaptive refinements and split-sample procedures that broaden the scope of applications. We also illustrated the methods across a wide range of model averaging schemes, including fixed weights, weights determined by OLS regressions, smoothed information criteria, the Mallows criterion, jackknife weighting, Bayesian model averaging, and bootstrap aggregation.

Taken together, these results show that valid interval prediction for model averaging is feasible even when the underlying models are misspecified and the data exhibit dependence. Beyond providing a practical tool for empirical work, the study also suggests open questions for conformal methods in dependent data settings, including extensions to more complex forms of serial correlation, as well as to high-dimensional model averaging.

\bibliographystyle{chicago}
\bibliography{ref.bib}

\end{document}

%% file: header.tex
    \usepackage[utf8]{inputenc}
    \usepackage[
        twoside,
        top=1in,
        bottom=0.75in,
        inner=0.5in,
        outer=0.5in
    ]{geometry}
    \usepackage{amsfonts,amsmath,amssymb}
    \usepackage{mathrsfs}
    \usepackage{enumerate}
    \usepackage{natbib}

\usepackage{algorithm,algorithmic}
\usepackage{tabularx}
\usepackage[figuresright]{rotating}
\usepackage{threeparttable}
\usepackage{subcaption,graphics}

\newcommand{\bx}{\mathbf{x}}

\newcommand{\bw}{\mathbf{w}}

\newcommand{\calC}{\mathcal{C}}

\def\CV{\mbox{CV}}

\newcommand{\F}{{\bf F}}

\newcommand{\w}{{\bf w}}

\newcommand{\x}{{\bf x}}

\newcommand{\y}{{\bf y}}

\newcommand{\f}{{\bf f}}

\newcommand{\e}{{\bf e}}

\newtheorem{theorem}{Theorem}